\documentclass[journal]{IEEEtran}

\usepackage[T1]{fontenc}
\usepackage{amsmath,amssymb,amsfonts,mathrsfs,bm}
\usepackage{amsthm}
\usepackage{cite}
\usepackage{array}
\usepackage[shortlabels]{enumitem}
\usepackage{graphicx}
\usepackage{epstopdf}
\usepackage{url}
\usepackage{color}
\usepackage{multirow}
\usepackage[table]{xcolor}
\usepackage[normalem]{ulem}
\usepackage{array}
\newcolumntype{L}[1]{>{\raggedright\let\newline\\\arraybackslash\hspace{0pt}}m{#1}}
\newcolumntype{C}[1]{>{\centering\let\newline\\\arraybackslash\hspace{0pt}}m{#1}}
\newcolumntype{R}[1]{>{\raggedleft\let\newline\\\arraybackslash\hspace{0pt}}m{#1}}
\usepackage{subcaption}
\usepackage{xparse}

\ifx\loadbreqn\undefined
	\relax
\else
	\usepackage{breqn} 
\fi



\usepackage{glossaries}
\newacronym{wrt}{w.r.t.}{with respect to}
\newacronym{iid}{i.i.d.}{independent and identically distributed}
\newacronym{MIMO}{MIMO}{mulitple-input multiple-output}
\newacronym{AOA}{AOA}{angle-of-arrival}
\newacronym{AOD}{AOD}{angle-of-departure}
\newacronym{LOS}{LOS}{line-of-sight}
\newacronym{NLOS}{NLOS}{non-line-of-sight}
\newacronym{TOA}{TOA}{time-of-arrival}
\newacronym{TDOA}{TDOA}{time-difference-of-arrival}
\newacronym{RSS}{RSS}{received signal strength}
\newacronym{GNSS}{GNSS}{Global Navigation Satellite System}

\usepackage{float}

\ifx\notloadhyperref\undefined
	\ifx\loadbibentry\undefined
		\usepackage[hidelinks,hypertexnames=false]{hyperref} 
	\else
		\usepackage{bibentry}
		\makeatletter\let\saved@bibitem\@bibitem\makeatother
		\usepackage[hidelinks,hypertexnames=false]{hyperref}
		\makeatletter\let\@bibitem\saved@bibitem\makeatother
	\fi
\else
	\relax
\fi

\usepackage[capitalize]{cleveref}
\crefname{equation}{}{}
\Crefname{equation}{}{}
\crefname{claim}{claim}{claims}
\crefname{step}{step}{steps}
\crefname{line}{line}{lines}
\crefname{dmath}{}{}
\crefname{dseries}{}{}
\crefname{dgroup}{}{}
\crefname{Theorem}{Theorem}{Theorems}
\crefname{Corollary}{Corollary}{Corollaries}
\crefname{Proposition}{Proposition}{Propositions}
\crefname{Lemma}{Lemma}{Lemmas}
\crefname{Definition}{Definition}{Definitions}
\crefname{Example}{Example}{Examples}
\crefname{Assumption}{Assumption}{Assumptions}
\crefname{Remark}{Remark}{Remarks}
\crefname{Rem}{Remark}{Remarks}
\crefname{remarks}{Remarks}{Remarks}
\crefname{Theorem_A}{Theorem}{Theorems}
\crefname{Corollary_A}{Corollary}{Corollaries}
\crefname{Proposition_A}{Proposition}{Propositions}
\crefname{Lemma_A}{Lemma}{Lemmas}
\crefname{Definition_A}{Definition}{Definitions}

\usepackage{algorithm,algorithmic}

\ifx\useTheoremCounter\undefined
\newtheorem{Theorem}{Theorem}
\newtheorem{Corollary}{Corollary}
\newtheorem{Proposition}{Proposition}

\else
\newtheorem{Theorem}{Theorem}

\newtheorem{Proposition}[theorem]{Proposition}
\fi

\newtheorem{Definition}{Definition}

\theoremstyle{remark}
\newtheorem{Rem}{Remark}

\newcommand{\calF}{\mathcal{F}}
\newcommand{\calG}{\mathcal{G}}

\newcommand{\calI}{\mathcal{I}}

\newcommand{\calP}{\mathcal{P}}
\newcommand{\calQ}{\mathcal{Q}}

\newcommand{\calS}{\mathcal{S}}

\newcommand{\calX}{\mathcal{X}}

\newcommand{\calZ}{\mathcal{Z}}

\newcommand{\bx}{\mathbf{x}}

\newcommand{\bz}{\mathbf{z}}

\DeclareSymbolFont{bsfletters}{OT1}{cmss}{bx}{n}
\DeclareSymbolFont{ssfletters}{OT1}{cmss}{m}{n}
\DeclareMathSymbol{\bsfGamma}{0}{bsfletters}{'000}
\DeclareMathSymbol{\ssfGamma}{0}{ssfletters}{'000}
\DeclareMathSymbol{\bsfDelta}{0}{bsfletters}{'001}
\DeclareMathSymbol{\ssfDelta}{0}{ssfletters}{'001}
\DeclareMathSymbol{\bsfTheta}{0}{bsfletters}{'002}
\DeclareMathSymbol{\ssfTheta}{0}{ssfletters}{'002}
\DeclareMathSymbol{\bsfLambda}{0}{bsfletters}{'003}
\DeclareMathSymbol{\ssfLambda}{0}{ssfletters}{'003}
\DeclareMathSymbol{\bsfXi}{0}{bsfletters}{'004}
\DeclareMathSymbol{\ssfXi}{0}{ssfletters}{'004}
\DeclareMathSymbol{\bsfPi}{0}{bsfletters}{'005}
\DeclareMathSymbol{\ssfPi}{0}{ssfletters}{'005}
\DeclareMathSymbol{\bsfSigma}{0}{bsfletters}{'006}
\DeclareMathSymbol{\ssfSigma}{0}{ssfletters}{'006}
\DeclareMathSymbol{\bsfUpsilon}{0}{bsfletters}{'007}
\DeclareMathSymbol{\ssfUpsilon}{0}{ssfletters}{'007}
\DeclareMathSymbol{\bsfPhi}{0}{bsfletters}{'010}
\DeclareMathSymbol{\ssfPhi}{0}{ssfletters}{'010}
\DeclareMathSymbol{\bsfPsi}{0}{bsfletters}{'011}
\DeclareMathSymbol{\ssfPsi}{0}{ssfletters}{'011}
\DeclareMathSymbol{\bsfOmega}{0}{bsfletters}{'012}
\DeclareMathSymbol{\ssfOmega}{0}{ssfletters}{'012}

\DeclareMathOperator*{\argmax}{arg\,max}
\DeclareMathOperator*{\argmin}{arg\,min}

\newcommand{\qednew}{\nobreak \ifvmode \relax \else
      \ifdim\lastskip<1.5em \hskip-\lastskip
      \hskip1.5em plus0em minus0.5em \fi \nobreak
      \vrule height0.75em width0.5em depth0.25em\fi}

\newcommand{\nn}{\nonumber\\}

\newcommand{\ofrac}[1]{{\frac{1}{#1}}}

\newcommand{\tc}[1]{^{(#1)}}

\newcommand{\norm}[1]{{\left\lVert{#1}\right\rVert}}

\newcommand{\cond}[2]{\left. {#1}\, \middle| \, {#2} \right.}

\DeclareDocumentCommand \P { g d() g } {%
	\IfNoValueTF {#3} 
	{%
		\IfNoValueTF {#1} 
		{%
			\IfNoValueTF {#2}
			{%
				\mathbb{P}%
			}%
			{%
				\mathbb{P}\left({#2}\right)%
			}%
		}%
		{%
			\IfNoValueTF {#2}
			{%
				\mathbb{P}_{#1}%
			}%
			{%
				\mathbb{P}_{#1}\left({#2}\right)%
			}%
		}%
	}%
	{%
		\IfNoValueTF {#1} 
		{%
			\mathbb{P}\left(\cond{#2}{#3}\right)%
		}%
		{%
			\mathbb{P}_{#1}\left(\cond{#2}{#3}\right)%
		}%
	}%
}

\DeclareDocumentCommand \E { g o g } {%
	\IfNoValueTF {#3} 
	{%
		\IfNoValueTF {#1} 
		{%
			\IfNoValueTF {#2}
			{%
				\mathbb{E}%
			}%
			{%
				\mathbb{E}\left[{#2}\right]%
			}%
		}%
		{%
			\IfNoValueTF {#2}
			{%
				\mathbb{E}_{#1}%
			}%
			{%
				\mathbb{E}_{#1}\left[{#2}\right]%
			}%
		}%
	}%
	{%
		\IfNoValueTF {#1} 
		{%
			\mathbb{E}\left[\cond{#2}{#3}\right]%
		}%
		{%
			\mathbb{E}_{#1}\left[\cond{#2}{#3}\right]%
		}%
	}%
}

\definecolor{gray90}{gray}{0.9}

\newcommand{\blue}[1]{{{\color{black} #1}}}

\newcommand{\msout}[1]{\text{\color{green} \sout{\ensuremath{#1}}}}
\newcommand{\del}[1]{{\color{green}\ifmmode \msout{#1}\else\sout{#1}\fi}}

\newcommand{\hide}[1]{}

\renewcommand{\figurename}{Fig.}
\newcommand{\figref}[1]{\figurename~\ref{#1}}
\graphicspath{{./Figures/}} 
\DeclareGraphicsExtensions{.eps}

\usepackage{glossaries}
\newacronym{i.i.d.}{i.i.d.}{independent and identically distributed}
\newacronym{MFD}{MFD}{most favorable distribution}
\newacronym{MFH}{MFH}{most favorable hypothesis}
\newacronym{w.r.t.}{w.r.t.}{with respect to}
\newacronym{LFD}{LFD}{least favorable distributions}
\newacronym{FHE}{FHE}{Fully Homomorphic Encryption}
\newacronym{PBPO}{PBPO}{person-by-person optimization}

\newcommand{\GN}{G^{\circ}} 
\newcommand{\GMF}{G_{\mathrm{MF}}}
\newcommand{\tcalQ}{\widetilde{\calQ}}

\makeatletter
\def\munderbar#1{\underline{\sbox\tw@{$#1$}\dp\tw@\z@\box\tw@}}
\makeatother

\begin{document}
\title{Decentralized Detection with Robust Information Privacy Protection}

\author{Meng~Sun and
        Wee~Peng~Tay,~\IEEEmembership{Senior~Member,~IEEE}
\thanks{This research is supported in part by the Singapore Ministry of Education Academic Research Fund Tier 1 grant 2017-T1-001-059 (RG20/17), Singapore Ministry of Education Academic Research Fund Tier 2 grant MOE2018-T2-2-019, and the NTU-NXP Intelligent Transport System Test-Bed Living Lab Fund S15-1105-RF-LLF from the Economic Development Board, Singapore.}
\thanks{The authors are with the School of Electrical and Electronic Engineering, Nanyang Technological University, Singapore, e-mails: MSUN002@e.ntu.edu.sg, wptay@ntu.edu.sg}
}

\maketitle
\begin{abstract}
We consider a decentralized detection network whose aim is to infer a public hypothesis of interest. However, the raw sensor observations also allow the fusion center to infer private hypotheses that we wish to protect. We consider the case where there are an uncountable number of private hypotheses belonging to an uncertainty set, and develop local privacy mappings at every sensor so that the sanitized sensor information minimizes the Bayes error of detecting the public hypothesis at the fusion center, while achieving information privacy for all private hypotheses. We introduce the concept of a most favorable hypothesis (MFH) and show how to find a MFH in the set of private hypotheses. By protecting the information privacy of the MFH, information privacy for every other private hypothesis is also achieved. We provide an iterative algorithm to find the optimal local privacy mappings, and derive some theoretical properties of these privacy mappings. Simulation results demonstrate that our proposed approach allows the fusion center to infer the public hypothesis with low error while protecting information privacy of all the private hypotheses.
\end{abstract}


\section{Introduction}\label{sec:Intro}
With a burgeoning number of IoT devices penetrating into all aspects of our lives\cite{ChanCampoEsteveEtAl2009,da2014internet,whitmore2015internet,shah2016survey,misra2016internet,li20185g}, privacy-related issues are attracting increasing interest \cite{zhou2017security,sivaraman2015network,perera2015big,
caron2016internet,hwang2015iot,liao2018hypothesis}. Users of these devices worry about being watched, listened to, or tracked by wearable devices and smart home appliances \cite{consumer_privacy}. Sensitive personal data like lifestyle preferences and location information may be abused for unwanted advertisement purposes or for more nefarious objectives like unauthorized surveillance. In China, for example, over $20$ million surveillance cameras equipped artificial intelligence have been installed. The new surveillance system have raised fears among citizens that the technology is being used to monitor their daily lives \cite{camera_watching}. To put consumers at ease and to encourage adoption of IoT technologies, privacy consideration should be incorporated into the core design of IoT solutions, with users being given more control over what information can be shared with the service providers.

\begin{figure}[!tb]
\centering
\includegraphics[width=0.45\textwidth]{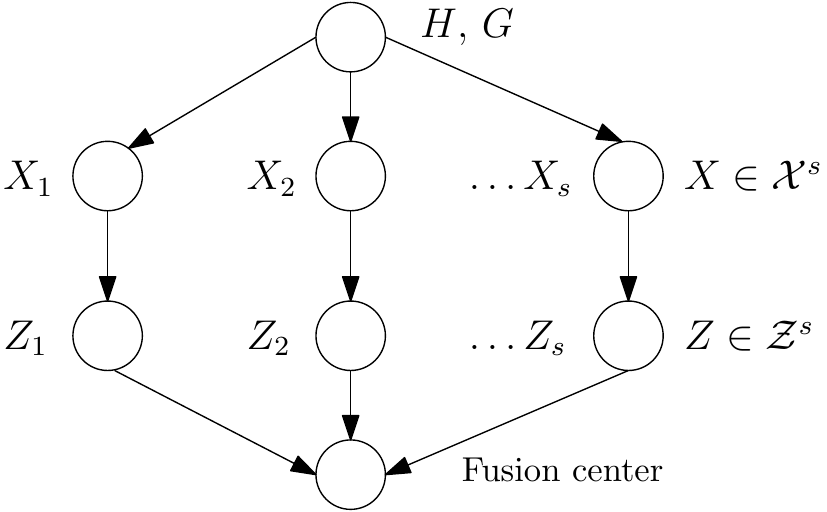}
\caption{An IoT network with public hypothesis $H$ and private hypothesis $G$. Each sensor $t$ makes a local observation $X_t$ that is mapped to $Z_t$, which is then sent to the fusion center.}
\label{fig:PBL}
\end{figure}

We model an IoT network consisting of multiple devices using the decentralized detection framework \cite{Tsi:93,Tsi:88,Tay:J12,SunTay:C16,Tay:J15,SunTay2017,SunTayHe2017towards} as shown in \cref{fig:PBL} (technical details are presented in \cref{sec:system model}). In this model, IoT devices or sensors make observations $X=(X_1,\ldots,X_s)$, where $X_t$ is the local observation of sensor $t \in \{1,\ldots,s\}$. Each sensor then maps its local observation $X_t$ to $Z_t$ and send that to a fusion center or service provider. The fusion center uses $Z=(Z_1,\ldots,Z_s)$ to infer a hypothesis of interest $H$, which is also called a public hypothesis. However, since the sensor information $Z$ may also be used by the fusion center to infer other hypotheses $G$ that the user may not have given authorization for (which are known as private hypotheses), sending unsanitized sensor information $Z=X$ to the fusion center may result in privacy leakage. For example, on-body wearables containing accelerometers and gyroscopes can be used to measure a person's movements and pose in order to detect falls. However, the sensor information may also be used to detect other activities performed by the person, leading to loss of privacy.

In this paper, we investigate the case where the fusion center is authorized to infer a public hypothesis $H$ based on sensor information that is sanitized in such a way that makes it difficult for the fusion center to infer an \emph{infinite set} of private hypotheses. To define the set of private hypotheses, we consider an uncertainty set \cite{Hub:65} defined around a nominal private hypothesis. This is because in most applications, one can define a specific public hypothesis, but it may be difficult for a user to specify multiple private hypotheses that she wants to protect. Our goal is to design \emph{local} privacy mappings $X_t \mapsto Z_t$ at each sensor $t$ to protect the information privacy \cite{PinCalmon2012} of private hypotheses that are ``close'' (in a specific sense as defined in \cref{sec:system model}) to the nominal hypothesis. We can interpret this framework as providing \emph{robust} privacy for the nominal private hypothesis, in the same spirit as robust hypothesis testing \cite{Hub:65,hub1973,moustakides1985robust,martin1971robust}.


\subsection{Related Work}
\label{subsec:related}
Various works \cite{AggarwalPhilip2008,asoodeh2016information,chechik2002extracting,Chen2014correlated,duchi2013local,dwork2008differential,diamantaras2016data,al2017ratio,al2017privacy,chaudhuri2011differentially,liao2017hypothesis,issa2017operational,zhao2018differentially} have investigated privacy preservation under different contexts, including privacy mechanisms for accessing databases and performing computations in cloud-based systems. All these works develop privacy techniques for centralized platforms where a trusted data curator is assumed to be available. In this paper, our aim is to develop privacy mechanisms for sensors in a decentralized network shown in \cref{fig:PBL}.

In a decentralized platform, without a trusted data curator, each sensor $t$ performs a local privacy mapping to sanitize its own data $X_t$ before sending a sanitized version $Z_t$ to a fusion center. The sanitized information $Z$ still enables the fusion center to detect $H$, while not disclosing private information $G$ to the fusion center. The papers \cite{wang2003using,xiong2016randomized,sarwate2014rate,liao2017hypothesis,imtiaz2018differentially,ren2018textsf} propose methods to achieve (local) differential privacy, while the works \cite{razeghiprivacy,mcsherry2009privacy,dwork2006our,duchi2013local} investigate privacy-preserving distributed data analytics. Other works resort to cryptographic techniques like \gls{FHE} \cite{Gentry2009}, partially homomorphic encryption \cite{zhang2018admm} and CryptoNets \cite{brakerski2014leveled,gilad2016cryptonets} to encrypt information from each sensor while allowing the fusion center to perform computations on the sensors' ciphertext. All the abovementioned methods aim at protecting data privacy, so that the raw data $X$ is not accessible by the fusion center. In this case, either statistical inference based on $Z$ of a private hypothesis is still feasible, or inference of the public hypothesis becomes severely degraded since data privacy does not distinguish between the public and private hypotheses \cite{SunTay2017}.

Information privacy as a metric to prevent statistical inference was first proposed by \cite{PinCalmon2012}. Information privacy ensures that the prior probability and posterior probability of a private hypothesis are similar. In \cite{SunTayHe2017towards,HeTaySun:C16,HeSunTay2018}, a nonparametric learning approach is proposed to determine the local sensor privacy mappings. The proposed approaches protect the information privacy of a single private hypothesis, while still allowing the fusion center to infer the public hypothesis with high accuracy. In this paper, we develop local privacy mappings for the sensors to protect the information privacy of an uncountable set of hypotheses close to a nominal private hypothesis. In contrast to \cite{SunTayHe2017towards,HeTaySun:C16,HeSunTay2018}, in this paper we consider parametric inference instead to quantify the utility-privacy trade-off. Furthermore, although the methods in \cite{SunTayHe2017towards,HeTaySun:C16,HeSunTay2018} can be extended to a finite set of private hypotheses, they cannot be easily generalized to handle an uncountable set of private hypotheses.

Robust detection was first proposed by \cite{Hub:65}. The paper \cite{hub1973} finds a general solution for the minimax test between two composite hypotheses. In \cite{moustakides1985robust}, robust decentralized detection is considered in an asymptotic setting, where \gls{i.i.d.} sensor observations is assumed. In all these works, a minimax test between a pair of representative simple hypotheses is conducted. In this paper, we study privacy-aware decentralized detection, in both non-asymptotic \cite{Tsi:93} and asymptotic \cite{Tsi:88} settings. Similar to the previous works, we use a representative pair of hypotheses when seeking robust privacy preservation for the private hypotheses. However, different from the previous works, where the main focus is to find a pair of distributions that is the \emph{hardest} to distinguish, the focus of our paper is to find a private hypothesis in the uncertainty set that is the \emph{easiest} to detect. The paper \cite{Kalantari2018} considers robust utility-privacy tradeoffs but does not use the term ``robust'' in the same context as this paper. Specifically, the authors do not consider an uncertainty set but instead characterize utility-privacy tradeoffs under different source distribution classes. 

\subsection{Our Contributions}
In this paper, we develop a privacy-preserving framework for a decentralized detection network. Our main contributions are as follows. 
\begin{enumerate}
\item  We introduce an uncertainty set to define an uncountable set of private hypotheses that are close in a specific sense to a nominal private hypothesis. We provide bounds for the Bayes error probability of detecting the public hypothesis.
\item We introduce the concept of a \emph{\gls{MFH}}, which is the hypothesis within the uncertainty set that is the easiest (in terms of the average Type I and II errors) to infer. By preserving the information privacy of the MFH, we achieve a guaranteed level of information privacy for all hypotheses within the uncertainty set. We show how to find a MFH.
\item We propose an iterative optimization algorithm to find the local privacy mappings at every sensor that achieve the optimal trade-off between utility (the Bayes error of detecting the public hypothesis) and information privacy for the set of private hypotheses. We show that each local privacy mapping is a randomization between at most two deterministic mappings. Under the asymptotic setting where the number of sensors goes to infinity, we show that the optimal sensor privacy mapping is a randomization between two deterministic sensor decision rules.
\end{enumerate}

The rest of this paper is organized as follows. In \cref{sec:system model}, we present our system model and introduce an uncertainty set to define the set of private hypotheses whose information privacy we wish to protect. In \cref{sec:error}, we present a bound on the utility and propose a series of relaxation steps. We also introduce and develop the concept of a MFH. In \cref{sec:method}, we propose an algorithm to solve for the optimal sensor privacy mappings. Simulation results are presented in \cref{sec:simulation}, and we conclude in \cref{sec:conclusion}.

\emph{Notations:}
We use capital letters like $X$ to denote random variables or vectors, lowercase letters like $x$ for deterministic scalars, and boldface lowercase letters like $\bx$ for deterministic vectors. The set $\Gamma^c$ denotes the complement of the set $\Gamma$. We use $p_X(\cdot)$ to denote the probability mass function of $X$, and $p_{X\mid Y}(\cdot \mid \cdot)$ to denote the conditional probability mass function of $X$ given $Y$. $\P$ denotes a generic probability measure, which will be clear from the context. We use $\norm{\cdot}_{TV}$ to denote total variation distance and $I(\cdot\ ;\ \cdot)$ to denote mutual information.

\section{Problem Formulation}
\label{sec:system model}

We consider a decentralized detection network as shown in \figref{fig:PBL}. The hypothesis $H \in\{0,1\}$ is the public hypothesis that the fusion center is authorized to infer, while $\GN \in \{0,1\}$ represents a private hypothesis that we wish to protect. Each sensor $t\in\{1,2,\ldots,s\}$ makes a noisy observation $X_t \in \calX=\{1,2,\ldots,|\calX| \}$, which is summarized using a randomized local decision rule or privacy mapping $q_t: \calX \mapsto \calZ=\{1,2,\ldots,|\calZ| \}$ to obtain $Z_t=q_t(X_t)$, before being sent to the fusion center. We interpret the privacy mapping $q_t$ as a conditional probability $q_t(z_t\mid x_t)=p_{Z_t\mid X_t}(z_t\mid x_t)$. We assume that conditioned on $X_t$, $Z_t$ is independent of everything else, and $\displaystyle\min_{h\in\{0,1\}} p_H(h) >0$. Based on the received $Z =(Z_t)_{t=1}^s$, the fusion center infers the public hypothesis $H$ using the Bayes detector $\gamma_H(Z)$. We denote $X = (X_t)_{t=1}^s$ and let $\calQ$ be the set of $p_{Z\mid X}$ such that
\begin{subequations}\label{Q}
\begin{align}
    &p_{Z\mid X}(\bz\mid \bx)=\prod_{t=1}^s q_t(z_t\mid x_t),\\
    &\sum_{z_t\in\calZ} q_t(z_t \mid  x_t) =1,\label{Q:sum}\\
    &q_t(z_t \mid x_t)\geq 0,\label{Q:0}\\
		&\sum_{x_t\in\calX}q_t(z_t|x_t)\geq\Delta,\label{Q:sum_pzx}\\
    &\forall\ x_t \in \calX,\ z_t\in\calZ,\ t=1,\ldots,s,\nonumber
\end{align}
\end{subequations}
where $\Delta\leq|\calX|/|\calZ|$ is a small positive constant. We impose the constraint in \cref{Q:sum_pzx} to ensure that every $z_t\in\calZ$ has a positive probability mass because otherwise a smaller $|\calZ|$ could have been used. 
Note that $\Delta$ has to be chosen no bigger than $|\calX|/|\calZ|$ since otherwise $|\calX|=\sum_{z_t,x_t}q_t(z_t|x_t) > |\calX|$, a contradiction.

\subsection{Uncertainty Set}
Given $p_{X,H,\GN}$, we define the utility to be the negative of the Bayes error $\P(\gamma_H(Z) \ne H)$ in inferring $H$ so that maximizing the utility over $\calQ$ means minimizing the Bayes error. Our goal is to find a $p_{Z\mid X} \in \calQ$ to achieve an optimal trade-off between the utility of inferring $H$ and protecting the privacy of $\GN$. However, in practice, it may be hard for a user to specify a particular $\GN$ to protect. Furthermore, there could be a model mismatch. For example, $\GN$ may correspond to a particular sensitive phenomenon but during the actual operation of the sensor network, a related but different phenomenon is realized instead. Therefore, instead of considering only $\GN$, we consider the privacy protection of all binary hypotheses $G$ that are close to $\GN$ in the following sense. We call $\GN$ the \emph{nominal} private hypothesis.

\begin{Definition}\label{def:uncertainty}
For a given $p_{X\mid\GN}$ and an uncertainty parameter $\delta \in [0,1)$, the uncertainty set $\calG(p_{X|\GN}, \delta) $ of private hypotheses \gls{w.r.t.} the nominal hypothesis $\GN$ is given by
\begin{align}\label{uncertainty set}
\begin{aligned}
&\calG(p_{X\mid\GN}, \delta) =\Big\{G\ : \min_{g\in\{0,1\}} p_G(g) >0,\\
&\quad\quad p_{X\mid G}(\bx\mid g)=(1-\delta)p_{X\mid  \GN}(\bx\mid g)+\delta f_g(\bx),\\
&\quad\quad f_g\in\calS_{\calX^s},\ g\in\{0,1\},\ \bx\in\calX^s\Big\},
\end{aligned}
\end{align}
where $\calS_{\calX^s}$ is the set of probability mass functions over $\calX^s$.
\end{Definition}

\begin{Rem}\label{rem:2-alternating}
\Cref{def:uncertainty} utilizes the concept of a contamination uncertainty class from the robust detection literature \cite{Hub:65,moustakides1985robust,martin1971robust}. A more general definition of an uncertainty class with desirable properties 
\blue{(e.g., the existence of a pair of least favorable distributions (LFDs) for the two uncertainty classes $\{p_{X\mid G}(\cdot\mid 0)\}$ and $\{p_{X\mid G}(\cdot\mid 1)\}$})
is the set of distributions upper bounded by a 2-alternating capacity \cite{hub1973}, which includes as special cases the contamination uncertainty set used in \cref{def:uncertainty}, and the uncertainty class consisting of $G$ such that $p_{X\mid G}$ is within a fixed total variation distance of $p_{X\mid \GN}$. Generalizing \cref{def:uncertainty} to utilize 2-alternating capacities is out of the scope of this paper, and is part of our future work. Using contamination uncertainty class is also easier to interpret in practice and leads to the following natural result in \cref{prop:uncertainty set}.
\end{Rem}

We assume that $\GN\in\calG(p_{X|\GN}, \delta)$. We note that $\calG(p_{X|\GN}, \delta)$ is an uncountable set if $\delta>0$. Abusing notation, to avoid clutter, we sometimes denote $\calG(p_{X|\GN}, \delta)$ as $\calG_X$ if $\delta$ and $\GN$ are clear from the context.

We also assume for some $\alpha>0$,
\begin{align}\label{asp:support_x}
p_{X|\GN}(\bx|g)\geq\alpha, \forall\ \bx\in\calX^s \text{ and } g\in\{0,1\}.
\end{align}
The assumption in \eqref{asp:support_x} is to avoid the trivial case where the nominal private hypothesis $\GN$ is perfectly detectable for some observations $\bx$.

In the example of fall detection using on-body wearables described in \cref{sec:Intro}, the public hypothesis $H$ is whether the person has fallen or not. The raw sensor information $X$ from the on-body wearables may be able to let the fusion center infer whether the person is performing a specific exercise like running, doing sit-ups, weight training, climbing stairs, etc. Suppose we define $\GN$ to be the hypothesis that the person is running at a particular constant speed $v$. Then, the uncertainty set $\calG(p_{X|\GN},\delta)$ consists of exercise hypotheses that are related to running from the sensors' perspective. To be more specific, depending on the value of $\delta$ chosen, the uncertainty set $\calG(p_{X|\GN}, \delta)$ can include hypotheses like the person is running at an average speed close to $v$, the person is running up a flight of stairs at a speed similar to $v$, or the person is performing jumping jacks at a rate that generates similar but ``noisier'' sensor readings $X$ than running at speed $v$. 

The motivation for our \cref{def:uncertainty} is to protect all hypotheses that generate similar sensor observations as the nominal hypothesis $\GN$. In this paper, for simplicity, we consider only a single private hypothesis uncertainty set, although our framework can be extended to include multiple uncertainty sets defined around multiple or $M$-ary nominal private hypotheses (see \cref{rem:Mary,{rem:multipleGN}} later).

\begin{Proposition}\label{prop:uncertainty set}
Let $g_*=\argmin_{g=0,1} p_{\GN}(g)$ and
\begin{align*}
\calF_\pi=\{G : p_{G\mid \GN}(0\mid 1)\leq \pi,\ p_{G\mid \GN}(1\mid 0)\leq \pi\},
\end{align*}
where
\begin{align*}
\pi<\min\{p_{\GN}(g_*),\min_{\bx,g}p_{X\mid \GN}(\bx\mid g)\}.
\end{align*}
Then $\calF_\pi \subset\calG(p_{X|\GN}, \delta)$, where
\begin{align}\label{t para}
\delta=1-\frac{p_{\GN}(g_*)-\pi}{(1-\pi)p_{\GN}(g_*)}\left(1-\frac{\pi}{\min_{\bx,g}p_{X\mid \GN}(\bx\mid g)}\right).
\end{align}
\end{Proposition}
\begin{IEEEproof}
See Appendix~\ref{prf:uncertainty set}.
\end{IEEEproof}
In \cref{prop:uncertainty set}, we can interpret $\calF_\pi$ to the set of outputs after passing $\GN$ through binary symmetric channels \cite[Chapter 7.1]{Cover2006} each of whose crossover probability is not more than $\pi$. Therefore, $G\in\calF_\pi \subset \calG(p_{X|\GN}, \delta)$ is a ``noisy'' version of $\GN$.

\subsection{Information Privacy and Optimization Problem Formulation}
Our goal is to prevent statistical inference of all $G\in\calG_X$. We choose information privacy \cite{PinCalmon2012} as the privacy metric because \cite{SunTay2017} has shown that the information privacy metric is one of the strongest privacy metrics that protects against statistical inference. The definition of information privacy is reproduced from \cite{PinCalmon2012} as follows.

\begin{Definition}[Information privacy]\label{def:info_privacy}
For $\epsilon > 0$, $G$ given $Z$ or $p_{G\mid Z}$ has $\epsilon$-information privacy, if for any $(g, \bz)$, we have
\begin{align}\label{ineq:info privacy}
    e^{-\epsilon}\leq\frac{p_{G \mid Z}( g \mid \bz)}{p_{G}(g)}\leq e^{\epsilon}.
\end{align}
The value $\epsilon$ is called the information privacy budget.
\end{Definition}

\Cref{def:info_privacy} essentially says that the received information $Z$ at the fusion center does not provide much information about $G$ in addition to what the fusion center already knows about it. Therefore, it is difficult to statistically infer $G$ given only $Z$. 
\blue{
\Cref{def:info_privacy} is similar to but different from the maximal realizable leakage defined in \cite[Definition~4]{issa2017operational}. The maximal realizable leakage is shown to be equivalent to $\max_{\bx,\bz} p_{Z\mid X}(\bz|\bx)/p_Z(\bz)$, which is related to the local differential privacy of the sensor data $X$ \cite{issa2017operational} and protects all possible hypotheses including $H$. This privacy notion is too strong for our purposes as we are interested in only protecting the privacy of those hypotheses closely related to the nominal private hypothesis $\GN$, while allowing the inference of $H$. 
}

Our optimization problem is formulated as follows: for a privacy budget $\epsilon > 0$, we aim to
\begin{subequations}\label{info_privacy original}
\begin{align}
& \min_{\gamma_H, p_{Z\mid X}\in\calQ} \P(\gamma_H(Z)\neq H)\label{obj_minH}\\
\text{s.t.}\ & e^{-\epsilon}\leq\frac{p_{Z\mid G}(\bz\mid g)}{p_{Z}(\bz)}\leq e^{\epsilon}, \label{original_pri}\\
&\forall\ G\in\calG(p_{X|\GN}, \delta),\ g\in\{0,1\},\ \bz\in\calZ^s,\nonumber
\end{align}
\end{subequations}
where we have made use of the equality $p_{G \mid Z}( g \mid \bz)/p_{G}(g)=p_{Z\mid G}(\bz\mid g)/p_{Z}(\bz)$. Note that here, $\P(\gamma_H(Z)\neq H)$ is the Bayes error \gls{w.r.t.} the probability distributions $p_{X,H,\GN}$ and $p_{Z\mid X}$.

\begin{Rem}\label{rem:utility}
The objective function in \cref{obj_minH} can be replaced by  
\begin{align*}
E(\gamma_H,p_{Z|X})
\end{align*}
where $E$ is a general loss function, examples of which are the following. Let $\P(\gamma(Z) \ne H; p_{X,H,G})$ denote the Bayes error for inferring $H$ \gls{w.r.t.} a given $p_{X,H,G}$, $p_{Z\mid X}$ and $\gamma$.
\begin{enumerate}[(a)]
\item\label{it:original_purpose} The formulation in \cref{obj_minH} uses $E=\P(\gamma(Z) \ne H; p_{X,H,\GN})$ where the Bayes error is \gls{w.r.t.} a given $p_{X,H,\GN}$.

\item\label{it:Bayesian_G} If a prior distribution over $G\in\calG(p_{X|\GN}, \delta)$ is known, a Bayesian formulation with $E=\E_G[P(\gamma(Z) \ne H; p_{X,H,G})]$, where the expectation is taken over all the private hypotheses $G\in\calG(p_{X|\GN}, \delta)$ can be used. 

\item\label{it:robust_H} A minimax or robust formulation involves assuming that each $p_{X,H,G}$ belongs to an uncertainty class 
\begin{align*}
\calP = \big\{p_{X,H,G} = p_{\GN}\cdot p_{X|G} \cdot p_{H\mid X,\GN} :\ \\
\quad\quad G\in \calG(p_{X\mid\GN}, \delta)\big\}
\end{align*}
and letting 
\begin{align*}
E=\max_{p_{X,H,G}\in\calP}\P(\gamma_H(Z)\neq H; p_{X,H,G}).
\end{align*}
It is easy to see that for any $p_{X,H,G}\in\calP$, $\bx\in\calX^s$, $h\in\{0,1\}$,
\begin{align}
&p_{X\mid H}(\bx | h) \nn
&\propto (1-\delta)\sum_g p_{X,H,\GN}(\bx,h,g) + \delta f(\bx,h), \label{pXH_uncertainty}
\end{align}
where $f(\bx,h)$ is some joint probability mass function for $(X,H)$. Then minimizing $E$ is equivalent to the robust detection of $H$ in \cite{Hub:65} over the contamination uncertainty classes defined by the class of distributions $\{p_{X\mid H}(\cdot \mid 0)\}$ and $\{p_{X\mid H}(\cdot\mid 1)\}$ of the form given in \cref{pXH_uncertainty} as $f(\bx,h)$ varies over all possible distributions. The loss function $E$ can then be simplified by finding the pair of LFDs of these two classes, which correspond to a $G'$ so that $E = \P(\gamma(Z)\ne H; p_{X,H,G'})$, which is similar to case~\ref{it:original_purpose}.
\end{enumerate}
Since the focus of this paper is on robust information privacy, we use case~\ref{it:original_purpose} for illustrative purposes throughout this paper.
\end{Rem}

\section{Error Bounds and Optimization Relaxation}\label{sec:error}

In this section, we first present bounds on the utility in the optimization problem \cref{info_privacy original}. We then propose a series of steps to relax \cref{info_privacy original}, and introduce the concept of a MFH, which simplifies our final optimization formulation significantly.

\subsection{Error Bounds}

\begin{Theorem}\label{thm:bound}
Suppose that $H$ has a uniform prior probability distribution under $p_{X,H,\GN}$. Then, the objective function in \eqref{info_privacy original} satisfies
\begin{align}
&\ofrac{2}-\frac{(e^\epsilon-1)\norm{p_{X\mid H}(\cdot\mid 0)-p_{X\mid H}(\cdot\mid 1)}_{TV}}{2\min_{G\in\calG_X} \{ \max\{m_G, e^\epsilon-1\}\}}\nonumber\\
&\geq  \min_{\gamma_H, p_{Z\mid X}\in\calQ} \P(\gamma_H(Z)\neq H)\nonumber\\
&\geq \ofrac{2}-\sqrt{\frac{I(H;X\mid \GN)+\epsilon}{2}},\label{utility_bound}
\end{align}
where $p_{X\mid H}(\bx\mid h) = \sum_{g=0,1} p_{X,\GN\mid H}(\bx, g\mid h)/2$, and for each $G\in\calG_X$ with corresponding joint distribution $p_{X,H,G}$,
\begin{align*}
&m_G = \max\Big\{|a_{G}(g)-b_{G}(g)|,|c_{G}(g)-d_{G}(g)| : g\in\{0,1\}\Big\},\\
&a_{G}(g)=\sum_{\bx\in I^+}\left(p_{X\mid G}(\bx\mid g)-e^{\epsilon}p_{X}(\bx; G)\right),\\
&b_{G}(g)=\sum_{\bx\in I^-}\left(p_{X\mid G}(\bx\mid g)-e^{\epsilon}p_{X}(\bx; G)\right),\\
&c_{G}(g)=\sum_{\bx\in I^+}\left(e^{-\epsilon}p_{X}(\bx; G)-p_{X\mid G}(\bx\mid g)\right),\\
&d_{G}(g)=\sum_{\bx\in I^-}\left(e^{-\epsilon}p_{X}(\bx; G)-p_{X\mid G}(\bx\mid g)\right),
\end{align*}
for $g=0,1$, and
\begin{align*}
& p_{X}(\bx; G) = \sum_{g=0,1} p_{X\mid G}(\bx\mid g)p_G(g),\\
&I^+=\{\bx\in\calX^s : p_{X\mid H}(\bx\mid 0)-p_{X\mid H}(\bx\mid 1)\geq 0\},\\
&I^-=\{\bx\in\calX^s : p_{X\mid H}(\bx\mid 0)-p_{X\mid H}(\bx\mid 1)<0\}.
\end{align*}
\end{Theorem}

\begin{IEEEproof}
See Appendix~\ref{prf:bound}.
\end{IEEEproof}

\begin{Rem}
From the proof of \cref{thm:bound}, the lower bound in \cref{utility_bound} can be generalized to case~\ref{it:Bayesian_G} in \cref{rem:utility} by replacing $I(H;X\mid\GN)$ (defined using $p_{X,H,\GN}$) with $I(H;X\mid G)$ (defined using $p_{X,H,G}$) and taking expectation \gls{w.r.t.} $G$. Similarly, it can be generalized to case~\ref{it:robust_H} in \cref{rem:utility} by replacing $I(H;X\mid\GN)$ with $I(H,X\mid G')$ (see \cref{rem:utility} for definition of $G'$). 
\end{Rem}

Both the bounds in \cref{thm:bound} approach to $1/2$ as $\epsilon\to 0$, which is expected as a budget $\epsilon=0$ means perfect information privacy. 
We present numerical results in \cref{sec:simulation} to show the behavior of these bounds for different $I(X;H)$. 

\subsection{Privacy Constraint Relaxation}\label{subsec:relaxation}

The optimization problem \eqref{info_privacy original} is a semi-infinite programming \cite[(1)]{lopez2007semi}. Without the constraints \cref{original_pri}, it is already a NP-complete problem \cite{TsiAth:85}. In the following, we present a series of relaxations so that a simple \gls{PBPO} approach can be developed (see \cref{sec:method}).

For any fusion rule $\gamma_G$ used to infer $G$ at the fusion center, let
\begin{align}
R_G(p_{Z\mid X},\gamma_{G})&=\ofrac{2}\Big(\P(\gamma_{G}(Z)=0){G=1} \nonumber\\
& \quad\quad +\P(\gamma_{G}(Z)=1){G=0}\Big), \label{R_G}
\end{align}
be the \emph{average of its Type I and Type II detection error probabilities} for $G$. Note that we are not assuming $p_G(0)=p_G(1)=1/2$. As will be clear from the following discussion and \cref{original_pri} itself, the prior distribution of $G$ does not affect the information privacy constraint in \cref{original_pri}. For any $\bz\in\calZ^s$ and hypothesis $G$, let
\begin{align*}
\ell_{Z\mid G}(\bz) &=\frac{p_{Z\mid G}(\bz\mid 1)}{p_{Z\mid G}(\bz\mid 0)}.
\end{align*}

From \cite[Proposition~1]{SunTayHe2017towards}, to achieve $\epsilon$-information privacy for $G\in\calG_X$, it suffices to ensure that
\begin{align}\label{X_pri}
\min_{\gamma_G} R_G(p_{Z\mid X},\gamma_{G})\geq \theta_G,
\end{align}
where $\theta_G = (1-c_G(1-e^{-\epsilon}))/2$ with
\begin{align*}
c_G=&\min\Big \{\P(Z\in\argmin_{\bz\in\calZ^s}\ell_{Z\mid G}(\bz)\mid G=0),\\
&\hspace{1.5cm}\P(Z\in\argmax_{\bz\in\calZ^s}\ell_{Z\mid G}(\bz)\mid G=1)\Big \}.
\end{align*}
\blue{The value $c_G$ corresponds to the minimum probability of the fusion center making the correct inference about $G$ over all realizations of $Z$. For example, if $Z=\bz$ minimizes $\ell_{Z\mid G}(\cdot)$, then the fusion center makes the correct inference that $G=0$.}
We have
\begin{align*}
\min_{G\in\calG_X}c_G
&\geq\min_{G\in\calG_X,\bz\in\calZ^s,g\in\{0,1\}}p_{Z|G}(\bz|g)\\
&=\min_{G,\bz,g}\sum_{\bx\in\calX^s}p_{Z|X}(\bz|\bx)p_{X|G}(\bx|g)\\
&\geq(1-\delta)\min_{\bz,g}\sum_{\bx\in\calX^s}p_{Z|X}(\bz|\bx)p_{X\mid\GN}(\bx|g)\\
&\geq(1-\delta)\alpha\Delta,
\end{align*}
where the penultimate inequality follows from the definition \cref{uncertainty set} and the last inequality is due to the assumptions \cref{Q:sum_pzx,asp:support_x}. Let
\begin{align}
\theta=\ofrac{2}\left(1-(1-e^{-\epsilon})(1-\delta)\alpha\Delta\right).\label{theta}
\end{align}
Then, $\theta\geq\theta_G$ for all $G\in\calG_X$. Therefore, by requiring that
\begin{align}\label{X_pri2}
\min_{G\in\calG_X,\gamma_{G}}R_G(p_{Z\mid X},\gamma_{G})\geq\theta,
\end{align}
the information privacy constraints \cref{original_pri} for all $G\in\calG_X$ are satisfied. 
To further simplify the constraint \cref{X_pri2}, we note that for any privacy mapping $p_{Z\mid X}\in\calQ$, 
\begin{align*}
\calG(p_{X|\GN}, \delta)\subset\calG_Z\triangleq\calG(p_{Z|\GN},\delta),
\end{align*}
since $p_{X\mid G}(\bx\mid g)=(1-\delta)p_{X\mid  \GN}(\bx\mid g)+\delta f_g(\bx)$ implies that 
\begin{align*}
p_{Z|G}(\bz|g)
&=\sum_{\bx\in\calX^s}p_{Z|X}(\bz|\bx)p_{X|G}(\bx|g)\\
&=(1-\delta)p_{Z\mid  \GN}(\bz\mid g)+\delta f'_g(\bz),
\end{align*}
where $f'_g(\bz)=\sum_{\bx\in\calX^s}p_{Z|X}(\bz|\bx) f_g(\bx) \in \calS_{\calZ^s}$. 


Therefore, our optimization problem \cref{info_privacy original} is relaxed to:
\begin{subequations}\label{info_privacy2}
\begin{align}
\min_{\gamma_H,p_{Z\mid X}\in\calQ}\ & \P(\gamma_H(Z)\neq H)\\
\text{s.t.}\ &\min_{G\in\calG_Z,\gamma_{G}}R_G(p_{Z\mid X},\gamma_{G})\geq\theta.\label{pri2}
\end{align}
\end{subequations}
It turns out that there exists a ``worst-case'' hypothesis in the set $\calG_Z$ so that achieving \cref{pri2} for this hypothesis implies information privacy for all other hypotheses in $\calG_Z$ with the same privacy budget. We discuss how to find this ``worst-case'' hypothesis in the next subsection.

\subsection{Most Favorable Hypothesis (MFH)}
\label{subsec:mfd}

In this subsection, we introduce the notion of a MFH \gls{w.r.t.} the privacy mapping $p_{Z\mid X}$. It is well known \cite[Section~II.B]{Poor2013} that the decision rule $\gamma_G^*$ minimizing \cref{R_G} over all $\gamma_G$ is given by
\begin{align*}
  \gamma_G^*(\bz)
	=\begin{cases}
    1,\text{ if  } \ell_{Z\mid G}(\bz) \geq 1,\\
    0, \text{ otherwise}.
   \end{cases}
\end{align*}
\blue{Recall that $R_G(p_{Z\mid X},\gamma_{G})$ is defined in \cref{R_G}.}

\begin{Definition}\label{def:mfd}
For a privacy mapping $p_{Z\mid X}\in\calQ$, if there exists a hypothesis $G\in\calG_Z$, such that
\begin{align*}
R_{G}(p_{Z\mid X}, \gamma_{G}^*)= \min_{G'\in\calG_Z}R_{G'}(p_{Z\mid X}, \gamma_{G'}^*),
\end{align*}
then we call $G$ a MFH \gls{w.r.t.}\ $p_{Z\mid X}$, and $p_{Z\mid G}$ a \gls{MFD} \gls{w.r.t.}\ $\calG_Z$.
\end{Definition}
The definitions of \gls{MFH} and \gls{MFD} mirror Huber's definition of a pair of \gls{LFD} \cite{Hub:65}. While the LFD is defined for a pair of uncertainty classes within a hypothesis, our \gls{MFH} and \gls{MFD} are defined for an uncertainty set of hypotheses. Similar to \cite{Hub:65}, we show below that a \gls{MFH} always exists and give an explicit construction of a \gls{MFD}.

\begin{Theorem}\label{thm:mfd}
For any privacy mapping $p_{Z\mid X}\in\calQ$ and $\calG_Z=\calG(p_{Z|\GN},\delta)$, we have
\begin{align} \label{eq:mfd_nominal}
\min_{G\in\calG_Z}R_{G}(p_{Z\mid X}, \gamma_G^*)=(1-\delta)R_{\GN}(p_{Z\mid X}, \gamma_{\GN}^*),
\end{align}
and a \gls{MFD} $p_{Z\mid  \GMF}$ is given by
\begin{subequations}\label{MFD}
\begin{align}
&p_{Z\mid  \GMF}(\bz\mid 0)\nonumber\\
&=\begin{cases}
        A_1(1-\delta)p_{Z\mid  \GN}(\bz\mid 1),\ &\bz=\munderbar{\bz} \\
        (1-\delta)p_{Z\mid  \GN}(\bz\mid 0),\ &\text{otherwise},
    \end{cases} \label{q0}
\end{align}
\begin{align}
&p_{Z\mid  \GMF}(\bz\mid 1)\nonumber\\
&=\begin{cases}
        A_2(1-\delta)p_{Z\mid  \GN}(\bz\mid 0),\ &\bz=\bar{\bz}\\
        (1-\delta)p_{Z\mid  \GN}(\bz\mid 1),\ &\text{otherwise},
    \end{cases}\label{q1}
\end{align}
\end{subequations}
where
\begin{align*}
\munderbar{\bz}&=\argmin_\bz \ell_{Z\mid \GN}(\bz),\\
\bar{\bz}&=\argmax_\bz \ell_{Z\mid \GN}(\bz),\\
A_1&=\frac{\frac{\delta}{1-\delta}+p_{Z\mid\GN}(\munderbar{\bz}\mid 0)}{p_{Z\mid\GN}(\munderbar{\bz}\mid 1)},\\
A_2&=\frac{\frac{\delta}{1-\delta}+p_{Z\mid\GN}(\bar{\bz}\mid 1)}{p_{Z\mid\GN}(\bar{\bz}\mid 0)},
\end{align*}
with $\munderbar{\bz}$ and $\bar{\bz}$ randomly chosen from the set of minimizers and maximizers, respectively, if they are not unique.
\end{Theorem}
\begin{IEEEproof}
See Appendix~\ref{prf:mfd}.
\end{IEEEproof}
\Cref{thm:mfd} provides a MFD w.r.t.\ $\calG_Z$, but the MFD may not be unique (consider for example the case where there are more than one $\bz$ that minimize $\ell_{Z\mid\GN}(\bz)$). Given a MFD $p_{Z\mid \GMF}$, the corresponding MFH $\GMF$ is also not unique (the MFH distribution $p_{\GMF}$ is any binary distribution satisfying the conditions in the uncertainty set definition in \cref{def:uncertainty} and $p_{X\mid\GMF}$ may not be uniquely determined by $p_{Z\mid\GMF}$). However, \cref{eq:mfd_nominal} always holds, which allows us to replace the constraint \cref{pri2} with a simpler constraint to obtain the following optimization problem:
\begin{subequations}\label{info_privacy3}
\begin{align}
\min_{\gamma_H,p_{Z\mid X}\in\calQ}\ & \P(\gamma_H(Z)\neq H) \label{util3}\\
\text{s.t.}\ &\min_{\gamma_\GN} R_{\GN}(p_{Z\mid X},\gamma_{\GN}) \geq \frac{\theta}{1-\delta}.\label{pri3}
\end{align}
\end{subequations}

\begin{Rem}\label{rem:Mary}
In this paper, we have assumed binary private hypotheses for simplicity. To generalize to $M$-ary private hypotheses, suppose that every $G\in\calG_X$ has range $\{0,1,\ldots,M-1\}$. For a detector $\gamma_{G,g}$ at the fusion center that distinguishes between the hypothesis pair $G=0$ and $G=g$, where $1\leq g \leq M-1$, let
\begin{align*}
R_{G,g}(p_{Z|X}, \gamma_{G,g}) &= \ofrac{2}\Big(\P(\gamma_{G,g}(Z)=g\mid G=0)\\
& \quad\quad +\P(\gamma_{G,g}(Z)=0\mid G=g)\Big).
\end{align*}
Then, from \cite[Theorem~2]{SunTayHe2017towards} and the discussions above, to achieve information privacy for all $G\in\calG_X$, it is sufficient to replace the privacy constraint \cref{pri3} by
\begin{align}\label{Mary_pri3}
\min_{1\leq g \leq M-1,\gamma_{\GN,g}} R_{\GN,g}(p_{Z|X}, \gamma_{\GN,g}) \geq \frac{\theta'}{1-\delta},
\end{align}
where $\theta'=\ofrac{2}\left(1-(1-e^{-\epsilon/2})(1-\delta)\alpha\Delta\right)$. In other words, the optimization problem \cref{info_privacy3} now has $M-1$ constraints. The privacy mapping design and results in the next \cref{sec:method} can then be generalized accordingly. 
\end{Rem}
\begin{Rem}\label{rem:multipleGN}
Similar to the case of $M$-ary private hypotheses, suppose that we have multiple nominal binary private hypotheses $(\GN_k)_{k=1}^K$, with corresponding uncertainty sets $(\calG(p_{X|\GN_k},\delta_k))_{k=1}^K$. Then, \cref{pri3} can be generalized to the $K$ constraints $\min_{\gamma} R_{\GN_k}(p_{Z\mid X},\gamma) \geq \theta/(1-\delta_k)$, for $k=1,\ldots,K$.
\end{Rem}

\section{Privacy Mapping Design}
\label{sec:method}

Since \cref{info_privacy3} is NP-complete (note that optimizing the objective is NP-complete \cite{TsiAth:85}), we propose a suboptimal approach in this section to obtain the privacy mapping $p_{Z\mid X}$. We also derive some properties of this privacy mapping. We consider both the non-asymptotic case and asymptotic case where the number of sensors $s\to\infty$.

\subsection{Fixed number of sensors}\label{subsec:PBPO}

\begin{algorithm*}[!tb]
\caption{\gls{PBPO} of local privacy mappings.}
\label{algo}
    \begin{algorithmic}[1]
    \STATE{\textbf{input:} $p_{X,H,\GN},\delta,r,\xi$}
    \STATE{\textbf{initialization:}\
        \begin{itemize}
        \item Let $p^{t,(0)}(z\mid x)=\ofrac{|\calZ|}+n_t(z\mid x),\forall z\in\calZ$, with $\sum_{z\in\calZ}n_t(z\mid x)=0,\forall x\in\calX,t=1,2,\ldots,s$.
        \item For fixed $p_{Z|X}^{(0)}=\prod_{t=1}^s q_t^{(0)}$, optimize $\theta_0$ in \eqref{threshold} with likelihood ratio test.
        \item Set $\theta=r\theta_0$.
        \item Set $k=-1$ and $E^{(0)}=1$.
        \end{itemize}
        }
    \REPEAT\label{repeat}
    \STATE{
        \begin{enumerate}[(i)]
            \item $k=k+1$.
            \item Fix $p_{Z\mid X}^{(k)}=\prod_{t=1}^s q_t^{(k)}$, optimize $\gamma_H^{(k+1)}$ to be the Bayes detector of hypothesis $H$.\label{algo_step:min_H}
            \item Fix $\gamma_H^{(k+1)}$ and $q_{\backslash t}^{ (k)}=\prod_{i=1}^{t-1} q_i^{(k+1)}\cdot\prod_{i=t+1}^s q_i^{(k)}$, for $t=1,2,\ldots,s$, solve \eqref{info_privacy_L} to find the optimal $q_t^{(k+1)}$  and the optimal solution $E^{(k)}$. The linear solver used is \cite[Algorithm 2]{Cohen2018}. \label{algo_step:min_S}

        \end{enumerate}
    }

    \UNTIL{
        $\frac{E^{(k)}-E^{(k+1)}}{E^{(k)}}\leq \xi$.
    }
    \RETURN $q_t=q_t^{(k+1)},t=1,2,\ldots,s$.

    \end{algorithmic}
\end{algorithm*}

We optimize \eqref{info_privacy3} using \gls{PBPO} as follows. For a fixed $p_{Z\mid X}$, we first find the Bayesian detector $\gamma_H$ that minimizes the error in detecting $H$. 
Let $q_{\backslash t}=\prod_{j\neq t} q_j$. Then for each sensor $t=1,\ldots,s$, we fix $\gamma_H$, $q_{\backslash t}$ and optimize \eqref{info_privacy3} over all privacy mappings $q_t$ satisfying \cref{Q:sum,Q:0,Q:sum_pzx}.  This process is iterated until a convergence criterion is satisfied. Let $\Phi$ be the set of \emph{deterministic} local privacy mappings (a conditional probability distribution $p_{Z_t\mid X_t}(\cdot\mid x_t)$ with all its mass at a particular $z_t\in\calZ$ for each $x_t\in\calX$). The set $\Phi$ is finite since $|\calX|\cdot|\calZ|<\infty$. Then, for a fixed $q_{\backslash t}$, we can write
\begin{align}\label{pbpoi}
p_{Z\mid X} = \sum_{\phi\in\Phi}\nu_\phi \phi\cdot q_{\backslash t},
\end{align}
where $\sum_{\phi\in\Phi} \nu_\phi=1$, with $\nu_\phi\geq 0$, for all $\phi\in\Phi$. Let $L_H(p_{Z\mid X}, \gamma_H)$ be $\P(\gamma_H(Z)\neq H)$ when the fusion rule for $H$ is $\gamma_H$ and the privacy mapping $p_{Z\mid X}$ is used. From \cref{pbpoi}, we have
\begin{align*}
L_H(p_{Z\mid X}, \gamma_H) = \sum_{\phi\in\Phi}\nu_\phi L_H(\phi\cdot q_{\backslash t},\gamma_H).
\end{align*}
Similarly,
\begin{align*}
\min_{\gamma_{\GN}} R_\GN(p_{Z\mid X}, \gamma_\GN)
&= \min_{\gamma_{\GN}} \sum_{\phi\in\Phi}\nu_\phi R_\GN( \phi\cdot q_{\backslash t},\gamma_\GN)\\
&\geq \sum_{\phi\in\Phi}\nu_\phi \min_{\gamma_{\GN}}R_{\GN}(\phi\cdot q_{\backslash t},\gamma_{\GN}).
\end{align*}
In our \gls{PBPO} procedure, when optimizing for each sensor $t$ with a fixed $\gamma_H$ and $q_{\backslash t}$, we solve the following relaxed optimization problem:
\begin{align}
\min_{\nu_\phi}\ &\sum_{\phi\in\Phi}\nu_\phi L_H(\phi\cdot q_{\backslash t},\gamma_H)\nonumber\\
\text{s.t.}\ & \sum_{\phi\in\Phi}\nu_\phi \min_{\gamma_{\GN}}R_{\GN}(\phi\cdot q_{\backslash t},\gamma_{\GN})\geq\frac{\theta}{1-\delta},\label{info_privacy_L}\\
&\sum_{\phi\in\Phi} \nu_\phi=1,\ \nu_\phi\geq 0,\ \forall\phi\in\Phi.\nonumber
\end{align}

\begin{Proposition}\label{prop:random}
The optimal solution of \cref{info_privacy_L} is a randomization between at most two deterministic privacy mappings.
\end{Proposition}
\begin{IEEEproof}
The optimization problem \cref{info_privacy_L} is a linear program, and has $|\calX|\cdot|\calZ|$ variables with $|\calX|\cdot|\calZ|+2$ constraints. From linear programming theory \cite{Tsi:88}, the number of constraints in the optimal solution for which equality holds is no smaller than the number of variables. Therefore, at most two of the weights $\{\nu_\phi: \phi\in\Phi\}$ are nonzero.
\end{IEEEproof}

\begin{Rem}\label{rem:theta}
In practice, it may not be easy to know \emph{a priori} what privacy budget $\epsilon$ to set. Therefore, instead of fixing a privacy budget $\epsilon$ and choosing $\theta$ as in \cref{theta}, we let $\theta=r\theta_0$, where $r\in(0,1)$ is called the \emph{privacy threshold ratio}, and 
\begin{align}
\theta_0
&=(1-\delta)\max_{p_{Z\mid X}\in\calQ}\min_{\gamma_{\GN}}R_{\GN}(p_{Z\mid X},\gamma_{\GN})\label{threshold}
\end{align}
corresponds to the maximal information privacy achievable for $\GN$ under the constraints imposed by $\calQ$. The maximization on the right hand side of \cref{threshold} is achieved by letting $q_t(z_t\mid x_t)=1/|\calZ|$, for all $z_t\in\calZ$, $x_t\in\calX$, and $t=1,\ldots,s$. With this privacy mapping, it can be shown that $Z$ and $\GN$ are independent, and 
\begin{align}\label{theta0}
\theta_0=\frac{1-\delta}{2}.
\end{align}
In some applications, we may include additional constraints in the privacy mapping set $\calQ$ in order to improve the utility. For example, we may wish to constrain $\left|q_t(z_t\mid x_t)-1/|\calZ|\right|\geq \Delta'$ for some $\Delta'>0$ to avoid a mapping that results in $Z_t$ being independent of $H$. In this case, $\theta_0$ can be found from \cref{threshold} using a PBPO approach. 
\end{Rem}

The above discussion on our proposed \gls{PBPO} approach is summarized in \cref{algo}. We initialize the privacy mapping for each sensor $t$ to be $q_t^{(0)}(z\mid x)=1/|\calZ|+n_t(z\mid x)$, where $n_t(z\mid x)$ is a small random noise, satisfying $\sum_{z\in\calZ}n_t(z\mid x)=0$, for all $x\in\calX$. This is done to help the PBPO procedure avoid local optimum points. For each iteration $k$, we perform the PBPO procedure for each sensor $t=1,\ldots,s$ in order. This is iterated until a convergence criterion is reached.

The optimization problem \cref{info_privacy_L} at each sensor $t$ is a linear program (LP), which can be solved efficiently in polynomial time using standard LP solvers. The storage complexity and time to encode this LP is $O(|\calZ|^{|\calX|})$ since each term in the sums needs to be computed. Its solution time complexity is at most $O(\left(|\calX||\calZ|\right)^{2.38}\log \xi_0^{-1})$, where $\xi_0$ is the relative accuracy of the LP solver (see \cite{Cohen2018} for a tighter bound), for each sensor at each iteration of \cref{algo}. 

\begin{Proposition}\label{prop:converge}
Algorithm~\ref{algo} converges to a critical point.
\end{Proposition}
\begin{IEEEproof}
In Algorithm~\ref{algo}, since the Bayes error probability of detecting $H$ is non-increasing in steps 4\ref{algo_step:min_H} and 4\ref{algo_step:min_S}, $E\tc{k}$ is non-increasing in each iteration. Since $E\tc{k}$ is lower bounded, $E^k$ converges. 
From \cite[Proposition 4]{Grippo2000}, the convergence of $E\tc{k}$ implies the convergence of $\gamma_H$ and $q_t, t=1,2,\ldots,s$ to limit points. By \cite[Proposition 5]{Grippo2000}, every limit point of $\gamma_H$ and $q_t, t=1,2,\ldots,s$ is a critical point for Algorithm~\ref{algo}. The proposition is now proved.
\end{IEEEproof}

\subsection{Large number of sensors}

In this subsection, we consider the case where the number of sensors goes to infinity. We assume that the sensor observations are conditionally \gls{i.i.d.}\ given the hypotheses, and every sensor $t$ uses an identical local privacy mapping $q_t=q$. We have $p_{Z\mid X}(\bz\mid\bx)=\prod_{t=1}^s q(z_t\mid x_t)$. 

Let $\tcalQ$ be the set of privacy mappings $q(z|x)$. For any $q\in\tcalQ$ such that $Z$, $H$ and $\GN$ are not independent, the error probabilities $\min_{\gamma_H} \P(\gamma_H(Z)\neq H) \to 0$ and $\min_{\gamma_\GN} R_{\GN}(p_{Z\mid X},\gamma_{\GN})\to 0$ exponentially fast as $s\to\infty$ \cite{Cover2006,Tsi:88}. In particular, we observe that other than using a trivial mapping that results in the worst case error probability for detecting $H$, there is no way to avoid violating the privacy constraint \cref{original_pri} for a fixed $\epsilon$ budget when $s$ is sufficiently large. This result follows from \cite[Proposition~1(i)]{SunTayHe2017towards} since $\min_{\gamma_\GN} R_{\GN}(p_{Z\mid X},\gamma_{\GN})\to 0$ as $s\to\infty$. Therefore, we consider replacing the objective and constraint in the optimization problem \cref{info_privacy3} with their respective error exponents. The error exponent of \cref{util3} is given by
\begin{align}
&\lim_{s\to\infty}-\ofrac{s}\log \min_{\gamma_H} \P(\gamma_H(Z)\neq H)\nonumber\\
& = C_H(p) := -\min_{\lambda\in[0,1]}\log\sum_{z\in\calZ}p_{Z_1\mid H}^\lambda(z\mid 1) p_{Z_1\mid H}^{1-\lambda}(z\mid 0),
\end{align}
while that for \cref{pri3} is
\begin{align}
&\lim_{s\to\infty}-\ofrac{s}\log \min_{\gamma_\GN} R_{\GN}(p_{Z\mid X},\gamma_{\GN})\nonumber\\
& = C_\GN(p) := -\min_{\lambda\in[0,1]}\log\sum_{z\in\calZ}p_{Z_1\mid \GN}^\lambda(z\mid 1) p_{Z_1\mid \GN}^{1-\lambda}(z\mid 0).
\end{align}
We then obtain the optimization problem
\begin{subequations}\label{infinite}
\begin{align}
\max_{p\in\tcalQ}\ & C_H(p)\\
\text{s.t.}\ & C_{\GN}(p) \leq \beta, \label{inf_pri}
\end{align}
\end{subequations}
for some $\beta > 0$. The constraint \cref{inf_pri} is a privacy \emph{rate} constraint that prevents the average error probability of inferring any $G\in\calG_X$ from decaying faster than a fixed rate $\beta$ as $s\to\infty$. This corresponds to preventing the information privacy leakage of any $G\in\calG_X$ from worsening faster than a predefined rate.

\begin{Proposition}\label{prop:calZ}
There is no loss in optimality in \cref{infinite} if we restrict $|\calZ| \leq |\calX|+1$.
\end{Proposition}
\begin{IEEEproof}
For any $p\in\tcalQ$, we have
\begin{align*}
e^{-C_H(p)}
&=\sum_{z\in\calZ}p_{Z_1}(z)d_H(z),\\
e^{-C_{\GN}(p)}&=\sum_{z\in\calZ}p_{Z_1}(z)d_\GN(z),
\end{align*}
where 
\begin{align*}
d_H(z)&=\frac{p_{H\mid Z_1}(0\mid z)}{p_H(0)}\left(\frac{p_{H\mid Z_1}(1\mid z)p_H(0)}{p_{H\mid Z_1}(0\mid z)p_H(1)}\right)^{\lambda_1},\\
d_{\GN}(z)&=\frac{p_{ \GN\mid Z_1}(0\mid z)}{p_{\GN}(0)}\left(\frac{p_{G\mid Z_1}(1\mid z)p_{\GN}(0)}{p_{ \GN\mid Z_1}(0\mid z)p_{\GN}(1)}\right)^{\lambda_2},\\
\lambda_1&=\argmin_{\lambda\in(0,1)}\log\sum_{z\in\calZ}p_{Z_1\mid H}^\lambda(z\mid 1) p_{Z_1\mid H}^{1-\lambda}(z\mid 0),\\
\lambda_2&=\argmin_{\lambda\in(0,1)}\log\sum_{z\in\calZ}p_{Z_1\mid  \GN}^\lambda(z\mid 1) p_{Z_1\mid  \GN}^{1-\lambda}(z\mid 0).
\end{align*}
For each $z\in\calZ$, let $w_z$ be the vector $(p_{X_1\mid Z_1}(x\mid z))_{x\in\calX}$. We can then write
\begin{align*}
\sum_{z\in\calZ} p_{Z_1}(z)[w_z,d_{H}(z),d_{\GN}(z)]=[p_{X_1},e^{-C_H(p)},e^{-C_{\GN}(p)}].
\end{align*}
The vector $[p_{X_1},e^{-C_H(p)},e^{-C_{\GN}(p)}]$ belongs to a space with dimension $|\calX|+1$. From Carath\'eodory's theorem \cite[Appendix~C]{el2011network},  since $w_z,d_H(z),d_{\GN}(z)$ are real-valued continuous functions of $p_{X_1|Z_1}(\cdot|z)$, we can find $Z'_1$ such that 
\begin{align*}
|\{z : p_{Z'_1}(z)\ne 0\}| \leq |\calX|+1,
\end{align*}
and 
\begin{align*}
\sum_{z\in\calZ} p_{Z'_1}(z)[w_z,d_{H}(z),d_{\GN}(z)]=[p_{X_1},e^{-C_H(p)},e^{-C_{\GN}(p)}].
\end{align*}
The proposition now follows.
\end{IEEEproof}

Recall that $\Phi$ is the set of deterministic local privacy mappings. Any $p\in\tcalQ$ can be written as a randomization of deterministic mappings or the convex combination $\sum_{\phi\in\Phi}\nu_\phi \phi$, where $\nu_\phi\geq 0$ for all $\phi\in\Phi$ and $\sum_{\phi}\nu_\phi=1$. We now consider the case where each sensor performs independent randomization over $\Phi$ \cite{Tsi:93}, and reports its choice of privacy mapping to the fusion center.

\begin{Proposition}\label{prop:randomization}
Suppose that each sensor performs independent randomization over $\Phi$. Then, the optimal solution of \eqref{infinite} is a randomization between at most two deterministic decision rules.
\end{Proposition}
\begin{IEEEproof}
From \cite[Lemma 1]{Tsi:88}, for $J\in\{H,\GN\}$, we have
\begin{align*}
\lim_{s\to\infty}\ofrac{s}\log\min\P(\gamma_J(Z)\neq J)
=-\min_{\lambda\in[0,1]}\sum_{\phi\in\Phi}\nu_\phi \mu_J(\phi,\lambda),
\end{align*}
where
\begin{align*}
\mu_J(\phi,\lambda)&=\log\sum_{z\in\calZ}u_J^\lambda(\phi,z,1) u_J^{1-\lambda}(\phi,z,0), \\
u_J(\phi,z,j) &= \sum_{x\in\phi^{-1}(z)} p_{X\mid J}(x\mid j).
\end{align*}

Therefore, \eqref{infinite} can be reformulated as
\begin{align}\label{infinite4}
\begin{aligned}
\min_{\nu_\phi, \lambda_1\in[0,1]}\ &\sum_{\phi\in\Phi}\nu_\phi \mu_H(\phi,\lambda_1)\\
\text{s.t.}\ &\min_{\lambda_2\in[0,1]} \sum_{\phi\in\Phi}\nu_\phi \mu_{\GN}(\phi,\lambda_2)\geq -\beta,\\
& \nu_\phi \geq 0,\forall \phi\in\Phi,\ \sum_{\phi\in\Phi}\nu_\phi=1.
\end{aligned}
\end{align}
For any fixed $\lambda_1$ and $\lambda_2$, the linear program \eqref{infinite4} has $|\calZ|\cdot|\calX|$ variables with $|\calZ|\cdot|\calX|+2$ constraints. From linear programming theory \cite{Tsi:88}, the number of constraints in the optimal solution for which equality holds is no smaller than the number of variables. Therefore, at most two of the $\nu_\phi$'s are nonzero, and the proposition is proved.
\end{IEEEproof}
\section{Simulations}\label{sec:simulation}

In this section, we carry out simulation to provide insights into how different parameters impact the performance of our proposed approaches. We first consider a network of $4$ sensors and a fusion center, with $\calX=\{1,2,\ldots,16\}$, and $\calZ=\{1,2\}$. 

\begin{figure}[!htb]
  \centering
  \includegraphics[width=0.5\textwidth]{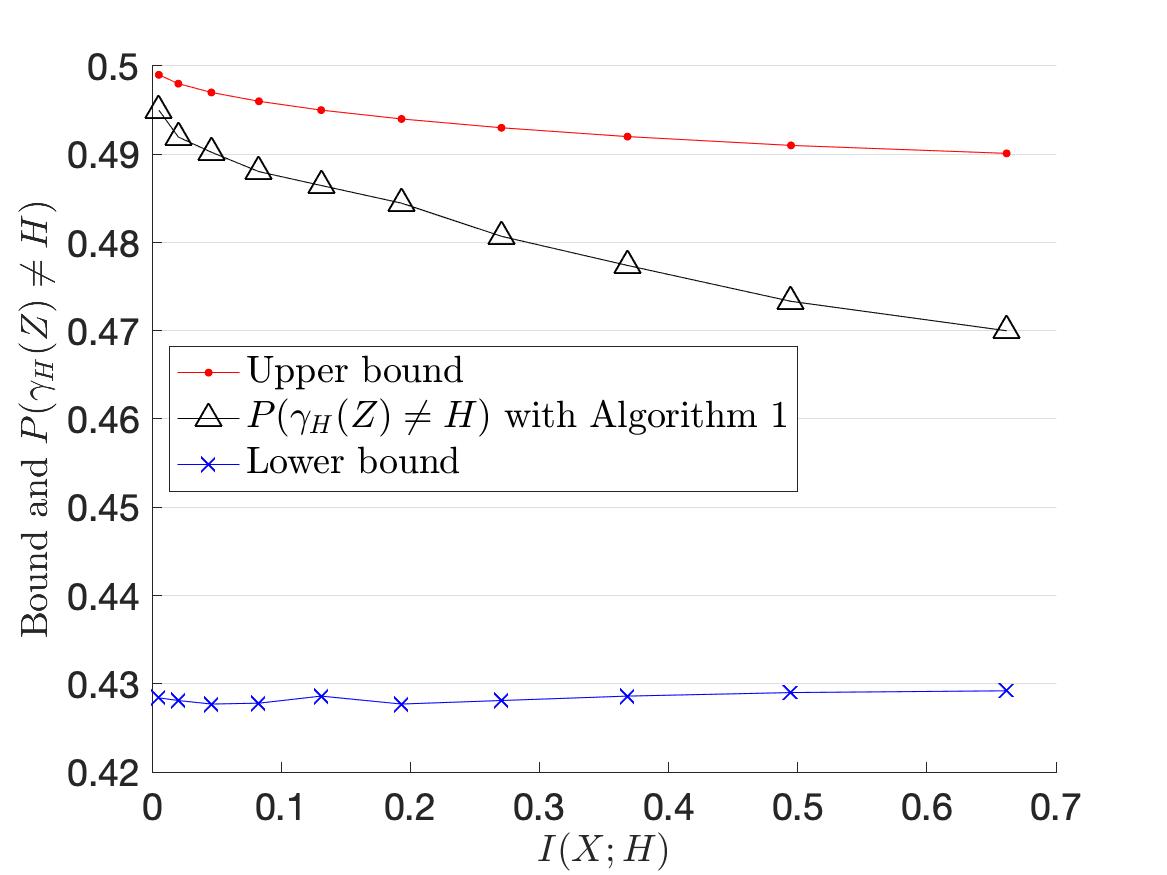}
  \caption{Upper and lower bounds in \cref{thm:bound} and $\P(\gamma_H(Z\neq H))$ calculated by Algorithm~\ref{algo} with varying $I(X;H)$.}
  \label{fig:bound}
\end{figure}

We generate different $p_{X,H,\GN}$ and calculate numerically the upper and lower bounds in \cref{thm:bound}. We set $I(X;H|\GN)=4\times 10^{-6}$ and impose an information privacy budget $\epsilon=0.01$. Different $p_{X,H,\GN}$ with different mutual information $I(X;H)$ are generated to test the performance of the bound. From \cref{fig:bound}, we see that the difference between the upper and lower bounds become tighter when $I(X;H)$ is large. Furthermore, the upper bound is tighter than the lower bound.

\begin{figure}[!htb]
  \centering
  \includegraphics[width=0.5\textwidth]{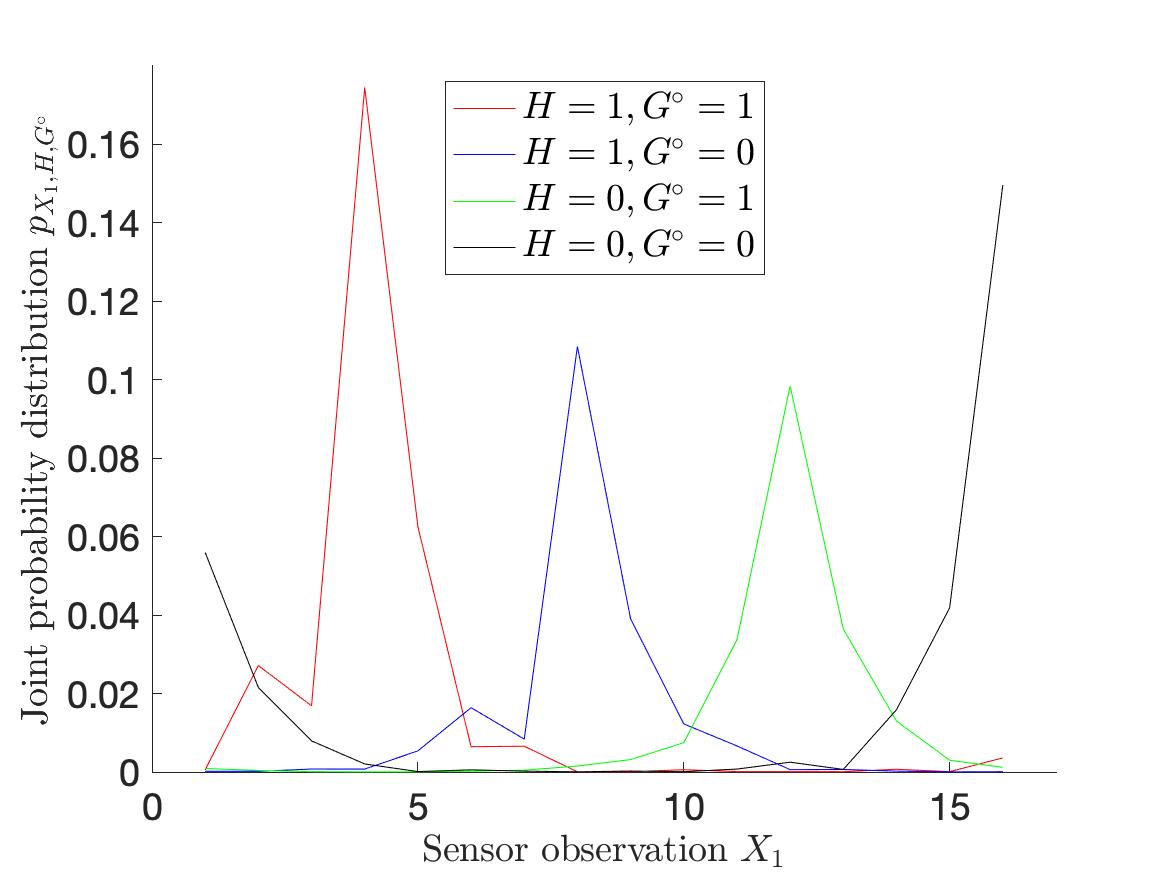}
  \caption{Joint distribution $p_{X_1,H,\GN}$  of sensor observation $X_1$, public hypothesis $H$, and private hypothesis $\GN$. The correlation coefficient between $H$ and $\GN$ is $0.2$.}
  \label{fig:distribution}
\end{figure}

We set the correlation coefficient between the public hypothesis $H$ and nominal private hypothesis $\GN$ to be $0.2$. The joint distribution $p_{X_1,H,\GN}$ at sensor 1 is shown in \figref{fig:distribution}. We assume that all sensors have the same joint distribution and are independent of each other conditioned on $(H,\GN)$. 

\begin{figure}[!htb]
  \centering
  \includegraphics[width=0.5\textwidth]{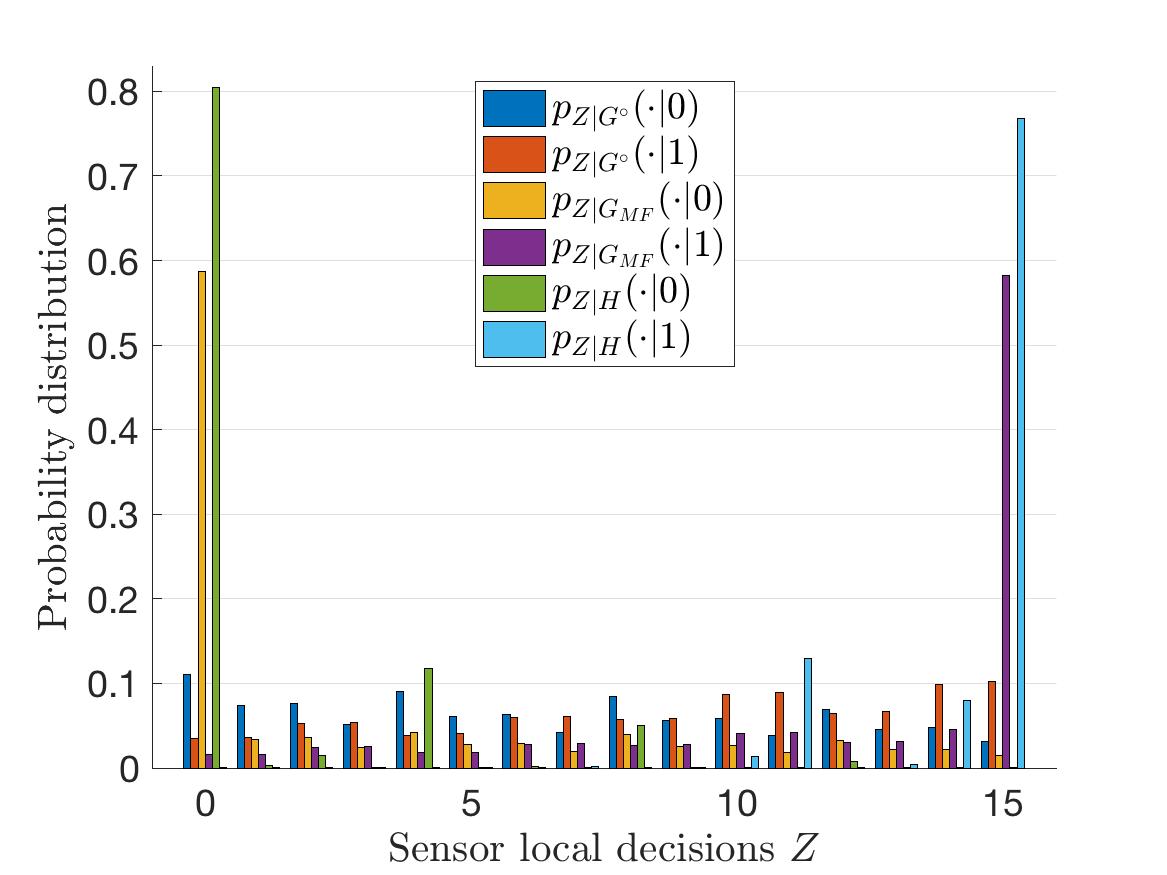}
  \caption{Conditional probability distribution of $Z=(Z_1,\ldots,Z_4)$ given different hypotheses. The horizontal axis shows the decimal form of the binary word $Z_1Z_2Z_3Z_4$.}
  \label{fig:Z}
\end{figure}

\begin{figure}[!htb]
  \centering
  \includegraphics[width=0.5\textwidth]{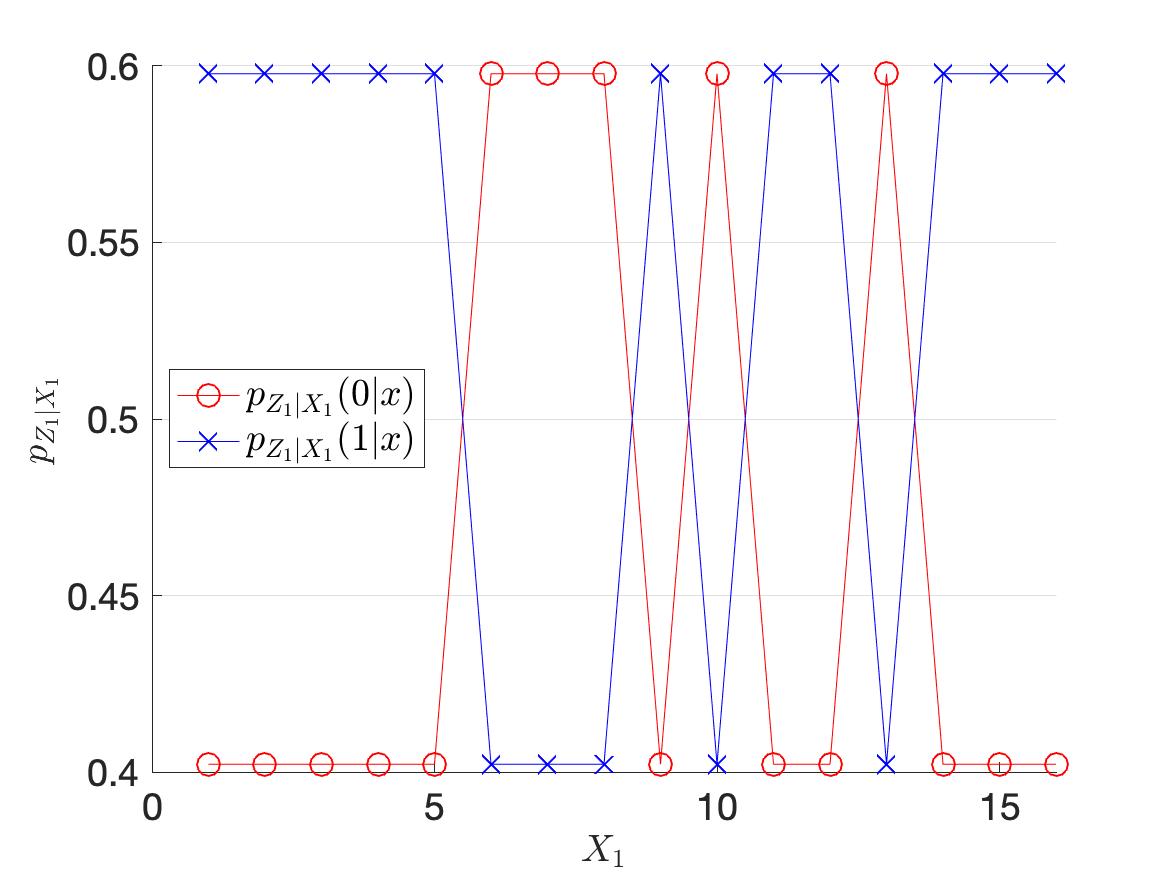}
  \caption{Privacy mapping at sensor 1, with $r=0.7$ and $\delta=0.54$.}
  \label{fig:Q}
\end{figure}

\begin{figure}[!htb]
\centering
	\begin{minipage}{0.45\textwidth}
  		\centering
		\includegraphics[width=\textwidth]{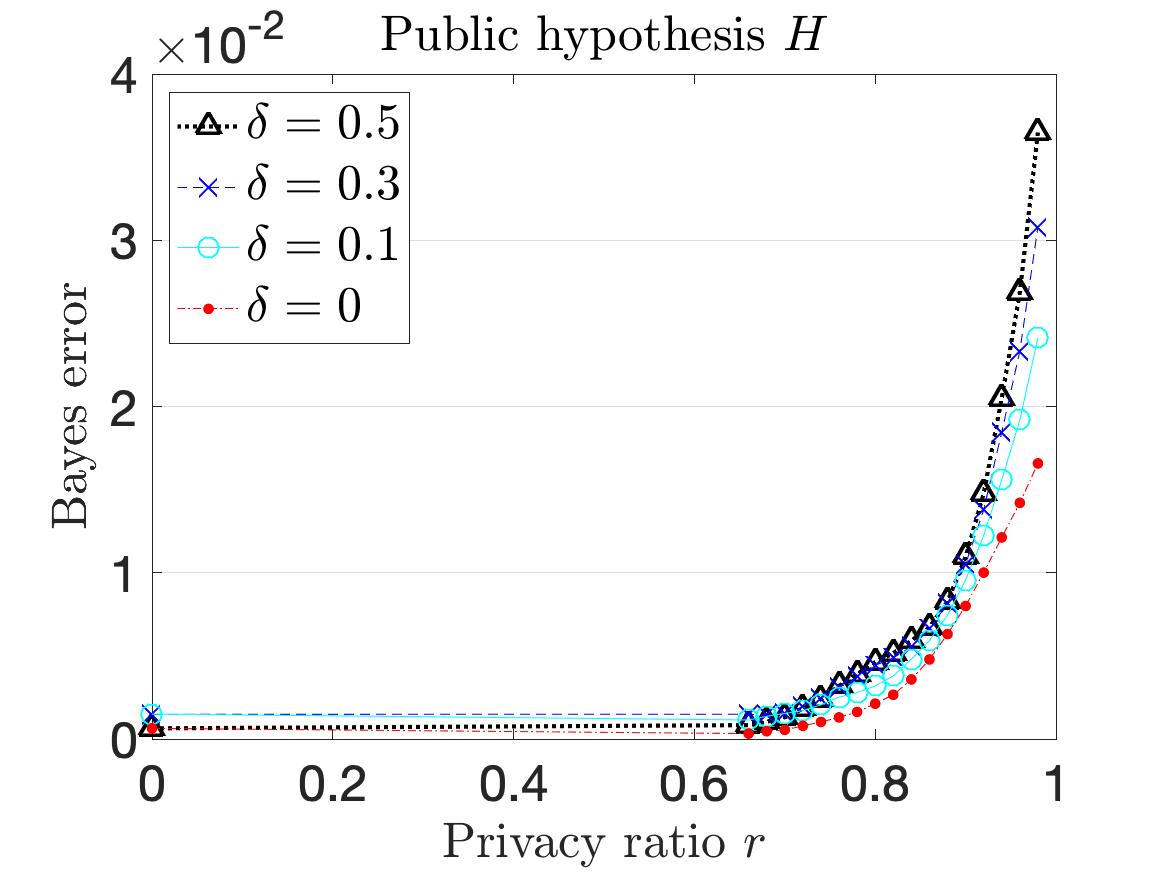}
		\subcaption{Detecting $H$.}\label{fig:ratio1}
	\end{minipage}%
	\vskip\baselineskip
	\centering
	\begin{minipage}{0.45\textwidth}
  		\centering
		\includegraphics[width=\textwidth]{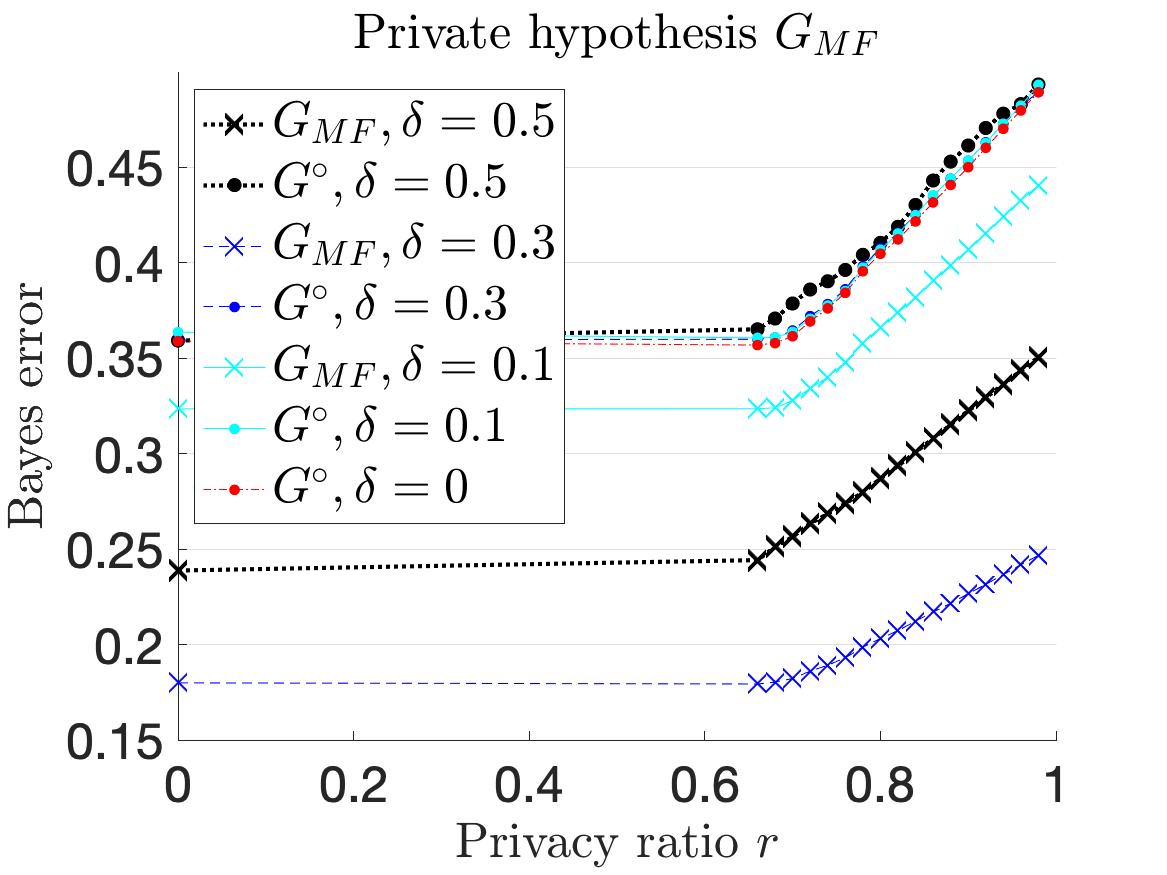}
		\subcaption{Detecting $G^\circ$ and $G_{\text{MF}}$}\label{fig:ratio2}
	\end{minipage}%
\caption{Bayes error probability of detecting $H$, $ \GN$ and $ \GMF$ with different privacy threshold ratio $r$ and uncertainty parameter $\delta$.} \label{fig:ratio}
\end{figure}

In our first simulation, we set the privacy threshold ratio $r=0.7$ and $\delta=0.54$ in \cref{algo} to obtain the privacy mapping $p_{Z\mid X}$. In \figref{fig:Z}, we visualize the distribution of $Z$ conditioned on different hypotheses, including the nominal private hypothesis $ \GN$, the MFH $\GMF$ (constructed using \cref{thm:mfd}), and the public hypothesis $H$. We observe that $p_{Z\mid\GN}(\cdot\mid0)$ and $p_{Z\mid\GN}(\cdot\mid1)$ are relatively similar to each other, which implies that $Z$ does not provide much statistical information about $\GN$. For the MFH $\GMF$, we have a similar observation that $p_{Z\mid\GMF}(\cdot\mid0)$ and $p_{Z\mid\GMF}(\cdot\mid1)$ are relatively similar to each other except at $Z=(0,0,0,0)$ and $Z=(1,1,1,1)$, which makes the MFH the easiest to distinguish amongst all hypotheses in $\calG_Z$. In this case, we find that the privacy budget $\epsilon=2.4$. We also observe that $p_{Z\mid H}(\cdot\mid0)$ and $p_{Z\mid H}(\cdot\mid1)$ are significantly different from each other, which implies that $H$ can still be inferred from $Z$ with good accuracy. We also show the privacy mapping at sensor $1$ in \cref{fig:Q} as an example.

In \figref{fig:ratio}, we show the Bayes error rates of detecting $H$ and $\GMF$ with different privacy threshold ratios $r$ and uncertainty parameters $\delta$. We let the privacy threshold $\theta=(1-\delta)r/2$ as given by \cref{theta0}. We see that the our proposed \cref{algo} yields privacy mappings that allow the fusion center to detect the public hypothesis $H$ with low error, while keeping the error for any private hypothesis $G\in\calG_X$ high. As expected, the Bayes error of detecting all hypotheses increases as the threshold ratio $r$ increases. Furthermore, the performance of our proposed \cref{algo} improves with smaller uncertainty parameter $\delta$. This is intuitive since a larger $\delta$ implies a stricter privacy constraint.

\begin{figure}[!htb]
  \centering
  \includegraphics[width=0.5\textwidth]{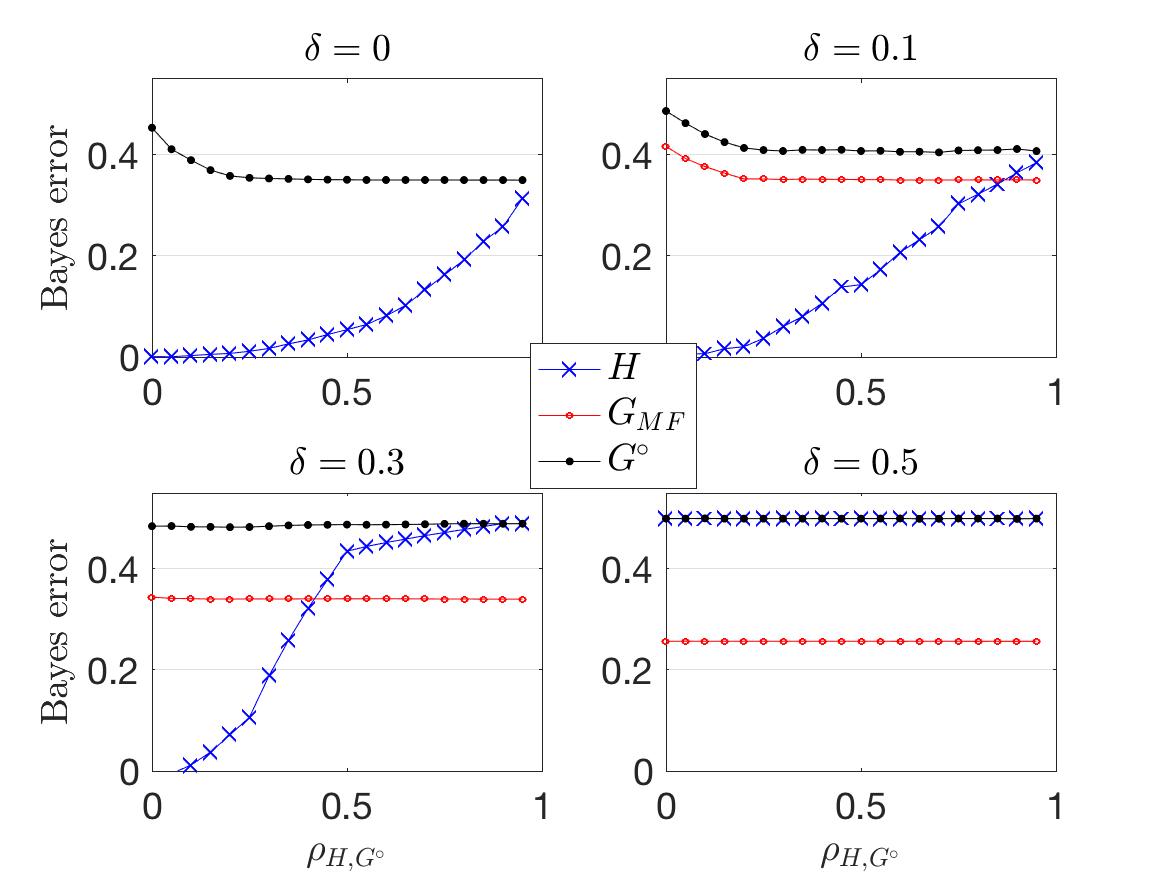}
  \caption{Bayes error of detecting $H$, $ \GN$ and $ \GMF$ with varying correlation coefficient between $H$ and $ \GN$. }
  \label{fig:cor}
\end{figure}

Let $\rho_{H,\GN}$ denote the correlation coefficient between $H$ and $\GN$. In \figref{fig:cor}, we fix $\theta=0.7$ and vary $\rho_{H,\GN}$. We see that if $\rho_{H,\GN}$ is small, the Bayes error of detecting $H$ is much smaller than the Bayes error of detecting any $G\in\calG_Z$. However, if $H$ and $ \GN$ are highly correlated, the Bayes error of detecting $H$ is close to that of $\GN$, and greater than that of detecting $\GMF$. This is because $\GMF$ is the \gls{MFH} in $\calG_Z$, and yields the smallest detection error amongst all $G\in\calG_Z$.

\subsection{Asymptotically Large Number of Sensors}

\begin{figure}[!htb]
  \centering
  \includegraphics[width=0.45\textwidth]{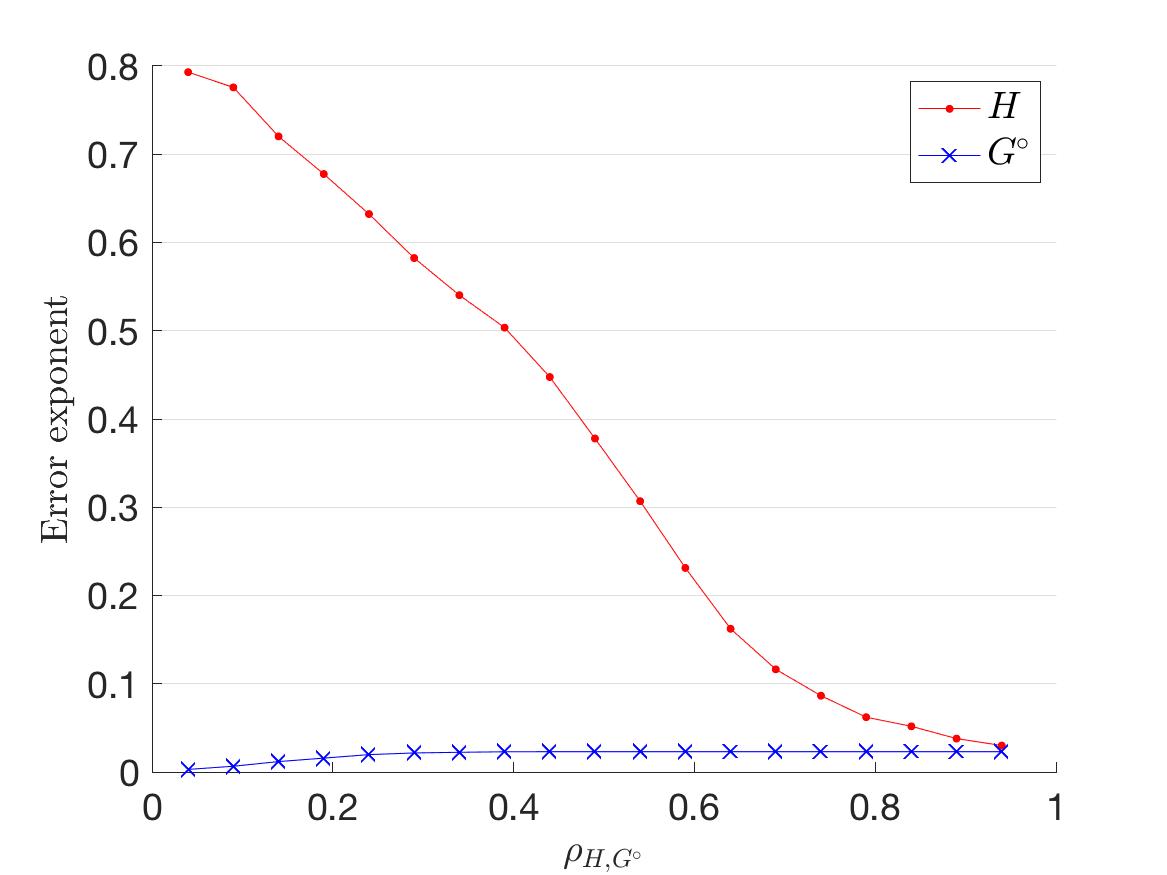}
  \caption{Error exponent for detecting $H$ and $\GN$ with varying correlation coefficient between $H$ and $\GN$.}
  \label{fig:infinite}
\end{figure}

Next, we plot the error exponents of detecting $H$ and $\GN$ as the number of sensors $s\to\infty$ in \figref{fig:infinite}. We use the same sensor observation distribution as in \cref{fig:distribution} and set $\beta=0.04$. We solve the optimization problem \cref{infinite} using \gls{PBPO} and \cref{prop:randomization} to obtain the optimal privacy mapping. We see that as the correlation coefficient between $H$ and $\GN$ increases, the error exponent of detecting $H$ decreases and approaches the error exponent of detecting $\GN$.

\begin{figure}[!htb]
  \centering
  \includegraphics[width=0.45\textwidth]{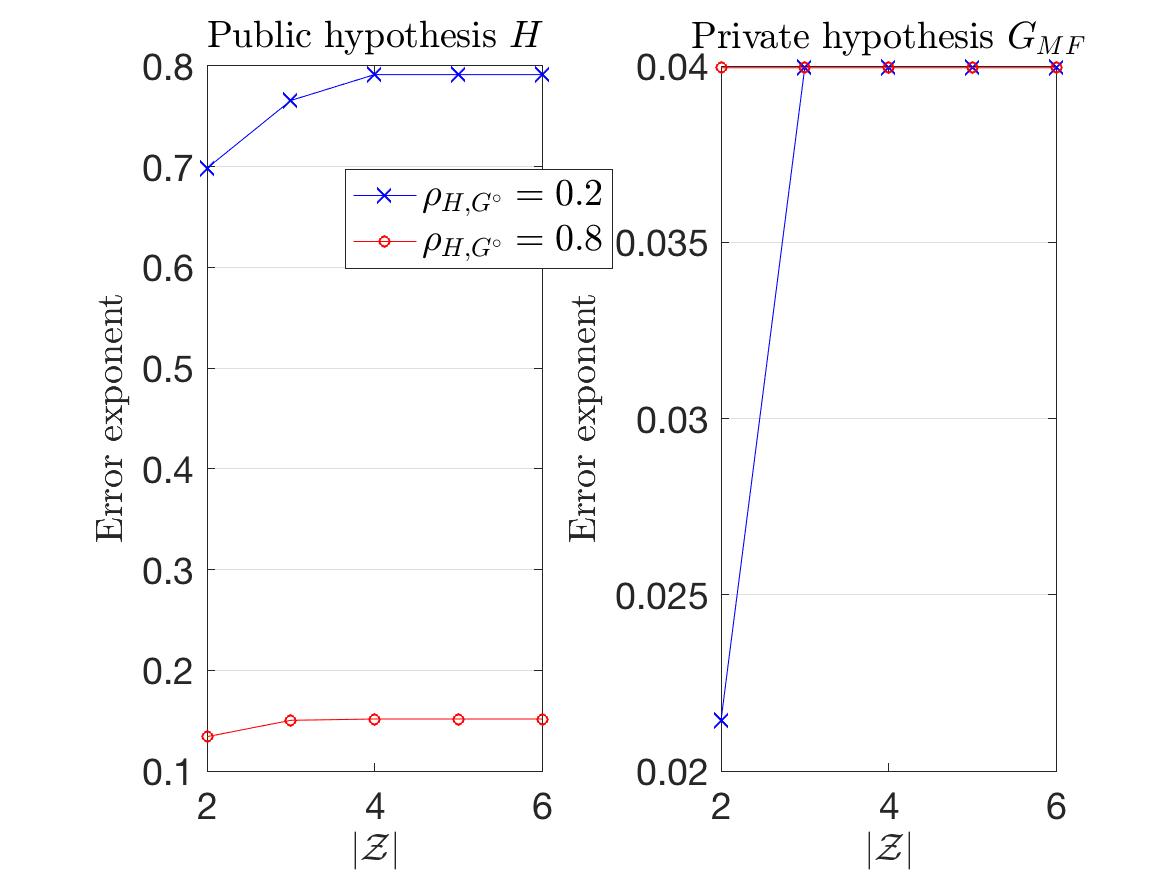}
  \caption{Error exponent for detecting $H$ and $\GN$ with varying $|\calZ|$. }
  \label{fig:card}
\end{figure}

In \figref{fig:card}, we study the accuracy-privacy tradeoff with varying $|\calZ|$. We set the sensor observation set cardinality $|\calX|=6$, and $\beta=0.05$. We consider the low-correlation case where $\rho_{H,\GN}=0.2$, and the high-correlation case where $\rho_{H,\GN}=0.8$. We see that as $|\calZ|$ increases, the error exponent of detecting $H$ increases and stays constant for $|\calZ|$ sufficiently large. The point where the error exponent becomes constant is larger for smaller correlation coefficient $\rho_{H, \GN}$. The results in \cref{fig:card} also verify \cref{prop:calZ}.

\subsection{Comparisons}\label{subsec:comparions}

Finally, we compare our proposed approach with approaches using other types of privacy metrics. We set $\rho_{H,\GN}=0.2$, and vary the distribution of $X$ so that the conditional mutual information $I(X;H|\GN)$ varies accordingly. We consider the following three privacy metrics for comparison. 

\begin{Definition}[Average information leakage \cite{PinCalmon2012}]\label{def:average}
Let $\epsilon_A\geq0$. We say that $\epsilon_A$-average information leakage is achieved if $I(\GN;Z)\leq\epsilon_A$.
\end{Definition}

\begin{Definition}[Maximal leakage \cite{liao2017hypothesis}]
Let $\epsilon_{ML}\geq 0$. We say that $\epsilon_{ML}$-maximal leakage privacy is achieved if $\sup_G\log\frac{\max_\gamma\P(\gamma(Z)=G)}{\max_{g}p_{G}(g)}\leq\epsilon_{ML}$. 
\end{Definition}

\begin{Definition}[Local differential privacy \cite{duchi2013local}]\label{def:local_diff}
Let $\epsilon_L\geq0$. We say that $\epsilon_L$-local differential privacy is achieved if for all neighboring $\bx,\bx'\in\calX^s$, and $\bz\in\calZ^s$, $\frac{p_{Z\mid X}(\bz\mid \bx)}{p_{Z\mid X}(\bz\mid \bx')}\leq e^{\epsilon_L}$.
\end{Definition}

\blue{
Average information leakage aims at limiting the mutual information between the private hypothesis $G$ and the sensor local decisions $Z$. Maximal leakage guarantees the Bayes error probability of detecting any private hypothesis to be large. Both privacy metrics prevent the fusion center from guessing the private hypothesis. Local differential privacy, on the other hand, prevents the fusion center from inferring the sensor observations ensuring that the sensor decision $Z=\bz$ and $Z=\bz'$ corresponding to neighboring sensor observations $X=\bx$ and $X=\bx'$ respectively are statistically similar. Here, $\bx$ and $\bx'$ are said to be neighboring observations if the two vectors differ at only one component.
}

The approaches we compare against when $\delta=0$ (i.e., $\calG_X = \{\GN\}$) are optimization formulations to achieve the minimum detection error while guaranteeing $\epsilon_A$-average information privacy and $\epsilon_L$-local differential privacy respectively. This ensures fair comparison as these approaches deal only with a single private hypothesis.  \blue{
We solve the optimization
problem (19) with the privacy constraint replaced by either the
average information leakage constraint or the maximal leakage
constraint, respectively. When $\delta>0$, we compare against the optimization formulation \cite[(3)]{liao2017hypothesis}
reproduced as follows: 
\begin{align*}
\max_{p_{Z|X}} \    & D\Big(\sum_{\bx\in\calX^s} p_{Z|X}(\bz|\bx)p_{X|G}(\bx|0)\\
&\hspace{1.5cm}||\sum_{\bx\in\calX^s}p_{Z|X}(\bz|\bx)p_{X|G}(\bx|1)\Big)\\
\text{s.t. \ } & \sum_{\bz\in\calZ^s}p_{Z|X}(\bz|\bx)=1,\forall \bx\in\calX^s \\
& p_{Z|X}(\bz|\bx)\geq 0, \forall \bz\in\calZ^s,\bx\in\calX^s,
\end{align*}
where $D(\cdot || \cdot)$  is the Kullback-Leibler divergence.
We approximate the problem in the high privacy domain, where the optimal $p_{Z|X}$ satisfies the following conditions \cite[Theorem~3]{liao2017hypothesis}:
\begin{enumerate}
\item The optimal $p_{Z|X}$ contains two unique columns. One of them has value $e^{\epsilon_{ML}}-1$ when $\bx\in\calI_+\overset{\Delta}{=}\{\bx:p_{X|G}(\bx|1)-p_{X|G}(\bx|0)\geq 0\}$. Another has value $e^{\epsilon_{ML}}-1$ when $\bx\in\calI_-\overset{\Delta}{=}\{\bx:p_{X|G}(\bx|1)-p_{X|G}(\bx|0)\leq 0\}$. 
\item Other columns have the same entry in each row, and each row sums to $1$.
\end{enumerate}
}
This optimization problem guarantees $\epsilon_{ML}$-maximal leakage privacy in the worst case private hypothesis.

\begin{figure}[!htb]
  \centering
  \includegraphics[width=0.45\textwidth]{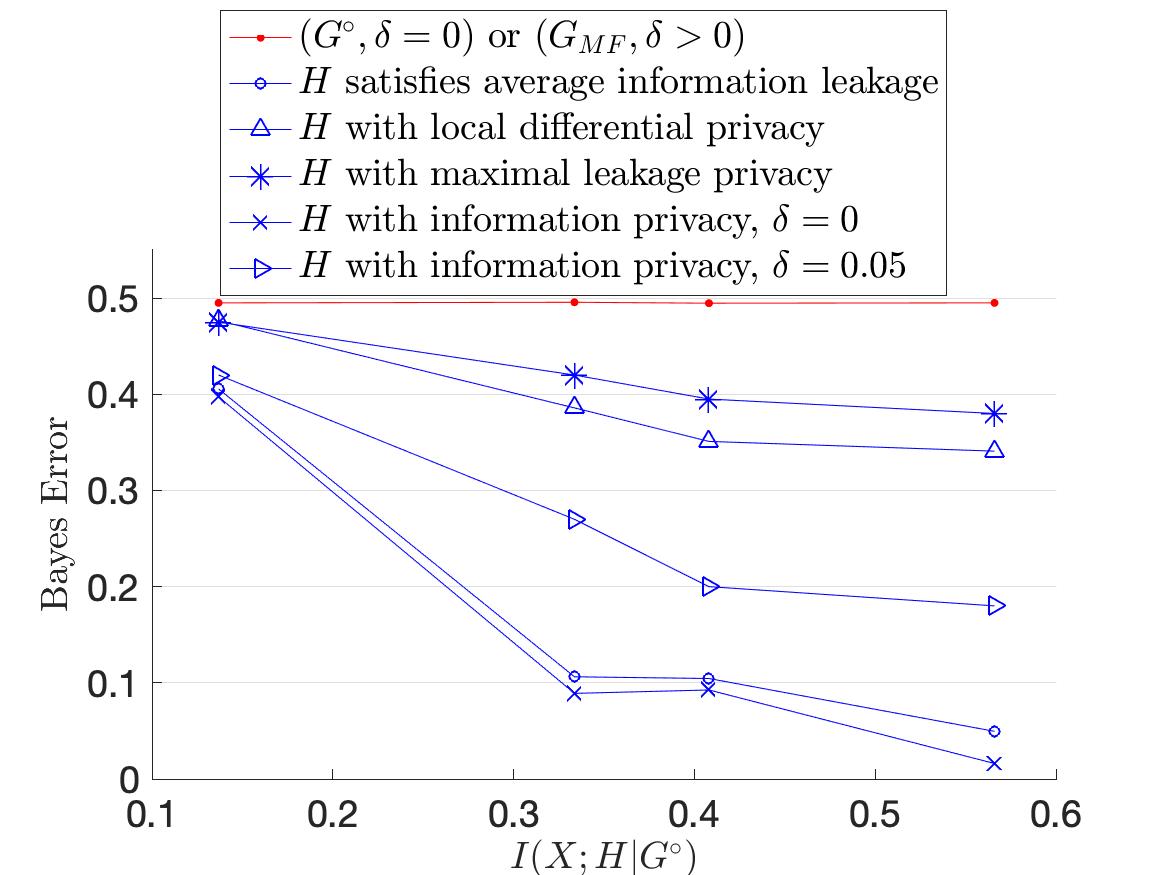}
  \caption{Bayes error for detecting $H$ and $\GN$ with varying conditional mutual information $I(X;H\mid \GN)$.}
  \label{fig:compare}
\end{figure}

In \figref{fig:compare}, we choose the privacy thresholds and budgets for the different metrics so that the Bayes error for detecting the private hypothesis $\GN$ is the same for all metrics when $\delta=0$. Similarly, when $\delta>0$, we set the Bayes error for detecting $\GMF$ to be the same across the metrics under comparison. We see that the Bayes error for detecting $H$ decreases as $I(X;H\mid\GN)$ increases. As expected, our proposed approach using information privacy as the privacy metric yields the minimum Bayes error for detecting $H$. By comparison, the approach using average information leakage has a slightly higher Bayes error for detecting $H$. The approach using local differential privacy has a significantly larger Bayes error for detecting $H$, since local differential privacy protects the \emph{data privacy} of the sensor observations $X$ and does not distinguish between statistical inferences for $H$ and $\GN$. For the $\delta >0$ comparison, the maximal leakage privacy formulation (3) in \cite{liao2017hypothesis} achieves the worst performance because it considers the strictest privacy criterion using the worst-case private hypothesis $G$. This serves as a benchmark upper bound for our utility performance if we let $\delta\to 1$ in our uncertainty set $\calG_X$.

\section{Conclusion}
\label{sec:conclusion}
In this paper, we have developed an approach to achieve information privacy for an uncertainty set of private hypotheses in a decentralized detection framework. We provided bounds for the utility of detecting the public hypothesis. We introduced the concept of a MFH of the uncertainty set of private hypotheses, and showed how to find a MFH, which allows us to transform the information privacy constraint into a more tractable metric that uses the average of the Type I and II error probabilities of detecting the MFH. We proposed a PBPO algorithm to find the local privacy mappings at each sensor. Simulations indicate that our approach produces higher error in detecting the private hypotheses, while maintaining a reasonably low error for detecting the public hypothesis. 

\blue{A future research direction is to consider our robust information privacy framework in the context of sequential sensor observations over time. Another interesting direction is to consider information privacy preservation in different network architectures like tree and tandem networks \cite{TanPatKle:91,TanPatKle:93,TayTsiWin:J08b,TayTsiWin:J08c,HoTayQue:J15}}. In this paper, we have assumed that the family of joint distributions between sensor observations and the hypotheses are known. In applications where such \emph{a priori} knowledge is not available, we can use a similar approach as in \cite{SunTayHe2017towards} to convert the objective \cref{util3} and privacy constraint \cref{pri3} into \emph{empirical risk} versions. Then, a nonparametric learning approach can be applied to learn the local sensor privacy mappings. An interesting future research direction is to characterize the utility-privacy tradeoff in this nonparametric learning framework.

\begin{appendices}
\section{Proof of Proposition~\ref{prop:uncertainty set}}
\label{prf:uncertainty set}
If $\pi=0$, the proposition trivially holds. We now assume that $\pi>0$ so that $0<\delta<1$ from \cref{t para}. Consider a $G\in\calF_\pi$. For any $g\in\{0,1\}$, let $\bar{g}=1-g$ and for any $\bx\in\calX^s$, let
\begin{align*}
f_g(\bx)=&\ofrac{\delta}(p_{X\mid G}(x\mid g)-(1-\delta)p_{X\mid \GN}(x\mid g)).
\end{align*}
To prove the proposition, it suffices to show that $f_g(\bx) \geq 0$ since $\sum_\bx f_g(\bx) = 1$. We have
\begin{align*}
p_{X\mid  \GN}(\bx\mid g)
&=\frac{p_{X, \GN,G}(\bx,g,g)}{\sum_{g'=0,1}p_{ \GN,G}(g,g')}+\frac{ p_{X, \GN,G}(\bx,g,\bar{g})}{p_{\GN}(g)}\\
&\leq p_{X\mid \GN,G}(\bx\mid g,g)+\frac{p_{X, \GN,G}(\bx,g,\bar{g})}{p_{\GN}(g)}\\
&\leq p_{X\mid \GN,G}(\bx\mid g,g)+p_{G\mid \GN}(\bar{g}\mid g),
\end{align*}
and
\begin{align}
p_{X\mid G}(\bx\mid g)
&=\sum_{g'=0,1}p_{X\mid \GN,G}(\bx\mid g',g)p_{ \GN\mid G}(g'\mid g) \nonumber\\
&\geq p_{ \GN\mid G}(g\mid g)(p_{X\mid \GN}(\bx\mid g)-p_{G\mid  \GN}(\bar{g}\mid g)).\label{pXGg}
\end{align}
We also have
\begin{align}
p_{\GN\mid G}(\bar{g}\mid g)
&= \frac{p_{G\mid \GN}(g\mid \bar{g})p_{\GN}(\bar{g})}{p_G(g)}\nonumber\\
&\leq \frac{\pi p_{\GN}(\bar{g})}{\sum_{g'=0,1} p_{G\mid \GN}(g\mid g')p_{\GN}(g')}\nonumber\\
&\leq \frac{\pi p_{\GN}(\bar{g})}{p_{G\mid \GN}(g\mid g)p_{\GN}(g)}\nonumber\\
&\leq\frac{\pi p_{\GN}(\bar{g})}{(1-\pi)p_{\GN}(g)}.\label{pGNG}
\end{align}
From \cref{pXGg,pGNG}, we obtain
\begin{align*}
&f_g(\bx)
=\ofrac{\delta}(p_{X\mid G}(\bx\mid g)-(1-\delta)p_{X\mid \GN}(\bx\mid g))\\
&\geq \ofrac{\delta}\Big(\frac{p_{\GN}(g)-\pi}{(1-\pi)p_{\GN}(g)}(p_{X\mid \GN}(\bx\mid g)-\pi)\\
& \hspace{5cm} -(1-\delta)p_{X\mid \GN}(\bx\mid g)\Big)\\
&= \frac{p_{X\mid \GN}(\bx\mid g)}{\delta}\Bigg(\frac{p_{\GN}(g)-\pi}{(1-\pi)p_{\GN}(g)}\left(1-\frac{\pi}{p_{X\mid \GN}(\bx\mid g)}\right)\\
& \hspace{7.5cm} -1+\delta\Bigg)\\
&\geq 0,
\end{align*}
where the last inequality follows from \cref{t para}. The proposition is now proved.

\section{Proof of Theorem~\ref{thm:bound}}
\label{prf:bound}
It can be shown that
\begin{align}
&\min_{\gamma_H,p_{Z\mid X}}\P(\gamma_H(Z)\neq H)\nonumber\\
&=\ofrac{2}-\ofrac{2}\max_{p_{Z\mid X}}\norm{p_{Z\mid H}(\cdot\mid 0)-p_{Z\mid H}(\cdot\mid 1)}_{TV}.\label{BayesH_TV}
\end{align}
We prove the theorem by bounding the total variation term in \cref{BayesH_TV}. We first prove the lower bound. From Pinsker's inequality \cite{csiszar2011information}, we have
\begin{align}
&\ofrac{2}\left(\sum_\bz |p_{Z\mid H}(\bz\mid 1)-p_Z(\bz)| \right)^2\nonumber\\
& \leq \sum_\bz p_{Z\mid H}(\bz\mid 1) \log \frac{p_{Z\mid H}(\bz\mid 1)}{p_Z(\bz)} \label{Pin1}
\end{align}
and
\begin{align}
&\ofrac{2}\left(\sum_\bz |p_{Z\mid H}(\bz\mid 0)-p_Z(\bz)| \right)^2 \nonumber\\
&\leq \sum_\bz p_{Z\mid H}(\bz\mid 0) \log \frac{p_{Z\mid H}(\bz\mid 0)}{p_Z(\bz)}. \label{Pin2}
\end{align}
From \cref{Pin1,Pin2}, we obtain
\begin{align}
&I(H;Z) \nonumber\\
\geq&
\frac{1}{4}\left(\sum_\bz \ofrac{2}|p_{Z\mid H}(\bz\mid 1)-p_{Z\mid H}(\bz\mid 0)| \right)^2 \nonumber\\
&  +\frac{1}{4}\left(\sum_\bz \ofrac{2}|p_{Z\mid H}(\bz\mid 1)-p_{Z\mid H}(\bz\mid 0)| \right)^2 \nonumber\\
=& \ofrac{2}\norm{p_{Z\mid H}(\cdot\mid 1)-p_{Z\mid H}(\cdot\mid 0)}_{TV}^2.\label{Ibound1}
\end{align}
We have
\begin{align}
I(H;Z)&=I(H;Z\mid \GN) + I(\GN;Z) - I(\GN;Z\mid H)\nonumber\\
&\leq I(H;X\mid \GN)+I(\GN;Z)\nonumber\\
&\leq I(H;X\mid \GN)+\epsilon,\label{Ibound2}
\end{align}
where the last inequality follows from $I(\GN;Z) = \sum_{\bz,g} p_{Z,\GN}(\bz,g)\log(p_{\GN\mid Z}(g\mid\bz)/p_\GN(g)) \leq \epsilon$ and \cref{original_pri}. From \cref{Ibound1,Ibound2}, we obtain
\begin{align*}
&\norm{p_{Z\mid H}(\cdot\mid 1)-p_{Z\mid H}(\cdot\mid 0)}_{TV}\\
&\leq \sqrt{2I(H;Z)}\\
&\leq \sqrt{2I(H;X\mid \GN)+\epsilon},
\end{align*}
and the lower bound in \cref{utility_bound} follows from \cref{BayesH_TV}. 

We next prove the upper bound. Fix a subset of $\Gamma\in\calZ^s$, and consider a $p_{Z\mid X}$ such that
\begin{align*}
p_{Z\mid X}(\bz\mid \bx)=
\begin{cases}
A, \ \bx\in I^+,\bz\in\Gamma\\
B, \ \bx\in I^-,\bz\in\Gamma
\end{cases},\\
p_{Z\mid X}(\bz\mid \bx)=
\begin{cases}
B, \ \bx\in I^+,\bz\in\Gamma^c\\
A, \ \bx\in I^-,\bz\in\Gamma^c
\end{cases}
\end{align*}
for some $A, B \geq 0$. We then have
\begin{align}
&\norm{p_{Z\mid H}(\cdot\mid 0)-p_{Z\mid H}(\cdot\mid 1)}_{TV}\nonumber\\
&=\ofrac{2}\sum_{\bz}|p_{Z\mid H}(\bz\mid 0)-p_{Z\mid H}(\bz\mid 1)| \\
&=\ofrac{2}\sum_{\bz}\left|\sum_{\bx}(p_{X\mid H}(\bx\mid 0)-p_{X\mid H}(\bx\mid 1))p_{Z\mid X}(\bz\mid \bx)\right| \nonumber\\
&=\ofrac{2}\sum_{\bz\in\Gamma}\bigg| A\sum_{\bx\in I^+}(p_{X\mid H}(\bx\mid 0)-p_{X\mid H}(\bx\mid 1))\nonumber\\
&\hspace{2cm} +B\sum_{\bx\in I^-}(p_{X\mid H}(\bx\mid 0)-p_{X\mid H}(\bx\mid 1))\bigg| \nonumber\\
&+\ofrac{2}\sum_{\bz\in\Gamma^c}\bigg| B\sum_{\bx\in I^+}(p_{X\mid H}(\bx\mid 0)-p_{X\mid H}(\bx\mid 1))\nonumber\\
&\hspace{2cm} +A\sum_{\bx\in I^-}(p_{X\mid H}(\bx\mid 0)-p_{X\mid H}(\bx\mid 1))\bigg| \nonumber\\
&=\ofrac{2}|\calZ|^s|A-B|\norm{p_{X\mid H}(\cdot\mid 0)-p_{X\mid H}(\cdot\mid 1)}_{TV},\label{eq:tv}
\end{align}
where \eqref{eq:tv} holds since
\begin{align*}
&\sum_{\bx\in I^+}(p_{X\mid H}(\bx\mid 0)-p_{X\mid H}(\bx\mid 1))\\
&=-\sum_{\bx\in I^-}(p_{X\mid H}(\bx\mid 0)-p_{X\mid H}(\bx\mid 1))\\
&=\norm{p_{X\mid H}(\cdot\mid 0)-p_{X\mid H}(\cdot\mid 1)}_{TV}.
\end{align*}
To obtain the upper bound in \cref{utility_bound}, we find $A\geq B$ that 
\begin{align*}
& \max_{A,B}\  A-B\\
& \text{s.t.}\  \\
& e^{-\epsilon}\leq\frac{A\sum_{\bx\in I^+}p_{X\mid G}(\bx\mid g)+B\sum_{\bx\in I^-}p_{X\mid G}(\bx\mid g)}{A\sum_{\bx\in I^+}p_{X}(\bx;G)+B\sum_{\bx\in I^-}p_{X}(\bx;G)}\leq e^{\epsilon},\\
& e^{-\epsilon}\leq\frac{B\sum_{\bx\in I^+}p_{X\mid G}(\bx\mid g)+A\sum_{\bx\in I^-}p_{X\mid G}(\bx\mid g)}{B\sum_{\bx\in I^+}p_{X}(\bx;G)+A\sum_{\bx\in I^-}p_{X}(\bx;G)}\leq e^{\epsilon},\\
& \text{for } g=0,1,\ \text{and all $G\in\calG_X$}, \\
&A\geq 0,B\geq 0,A+B=\frac{2}{|\calZ|^s}.  
\end{align*}
If we fix a $G\in\calG_X$, and let $A_G$ and $B_G$ be the corresponding $A$ and $B$ of the above optimization problem, we obtain the linear program:
\begin{align*}
\max_{A_G,B_G}\ & A_G-B_G\\
\text{s.t.}\ & a_G(g)A_G+ b_G(g)B_G\leq 0,\\
&b_{G}(g)A_G+a_{G}(g) B_G\leq 0,\\
& c_{G}(g)A_G+d_{G}(g) B_G\leq 0,\\
&d_{G}(g)A_G+c_{G}(g) B_G\leq 0,\\
& \hspace{2cm}\text{for}\ g=0,1,\\
&A_G\geq 0, B_G\geq 0,\\
&A_G+B_G=\frac{2}{|\calZ|^s},
\end{align*}
whose solution can be shown to be the following: Let 
\begin{align*}
F=\{(a_G(g),&b_G(g)),\ (b_G(g),a_G(g)),\\
 &(c_G(g),d_G(g)),\ (d_G(g),c_G(g)),\ g=0,1\}.
\end{align*}
\begin{enumerate}
\item If there exists a pair $(f_1,f_2)\in F$ such that $f_1>0$ and $f_2<0$, then we find a pair $(f^*_1,f^*_2)$ from $F$ such that $f^*_1> 0$ and $f^*_2< 0$ and that minimizes $f^*_1/f^*_2$. We obtain
\begin{align*}
A_G&=\frac{-2f^*_2}{|\calZ|^s(f^*_1-f^*_2)},\\
B_G&=\frac{2f^*_1}{|\calZ|^s(f^*_1-f^*_2)},
\end{align*}
and therefore $A_G-B_G=\frac{2(e^\epsilon-1)}{|\calZ|^s(f^*_1-f^*_2)}$, which is equivalent to
\begin{align*}
&A_G-B_G\\
&=\frac{2(e^\epsilon-1)}{|\calZ|^s\max\{|a_{G}(g)-b_{G}(g)|,|c_{G}(g)-d_{G}(g)|: g\in\{0,1\}\}}.
\end{align*}
\item If there does not exist any pairs $(f_1,f_2)\in F$ such that $f_1>0$ and $f_2<0$, we let $A_G=\frac{2}{\mid \calZ\mid ^s}$ and $B_G=0$ to obtain $A_G-B_G=\frac{2}{\mid \calZ\mid ^s}$.
\end{enumerate}
From \cref{eq:tv} and combining the two cases above, we obtain 
\begin{align*}
&\norm{p_{Z\mid H}(\cdot\mid 0)-p_{Z\mid H}(\cdot\mid 1)}_{TV} \\
& \geq \frac{(e^\epsilon-1)\norm{p_{X\mid H}(\cdot\mid 0)-p_{X\mid H}(\cdot\mid 1)}_{TV}}{\max\{m_G, e^\epsilon-1\}},
\end{align*}
for any $G\in\calG_X$. Maximizing over all $G\in\calG_X$, we obtain the upper bound in \cref{utility_bound}. The proof of the theorem is now complete.

\section{Proof of Theorem~\ref{thm:mfd}}
\label{prf:mfd}

If $\delta=0$, the theorem trivially holds. Suppose that $0<\delta<1$. It is easy to check that \eqref{MFD} defines a probability distribution in $\calG_Z$. For any $G\in\calG_Z$, let
\begin{align*}
\Gamma_G=\{\bz : \ell_{Z\mid G}(\bz) \geq 1\}.
\end{align*}
We have $\ell_{Z\mid \GN}(\munderbar{\bz}) \leq 1 \leq \ell_{Z\mid \GN}(\bar{\bz})$, and
\begin{align*}
A_1>\frac{p_{Z\mid\GN}(\munderbar{\bz}\mid 0)}{p_{Z\mid\GN}(\munderbar{\bz}\mid 1)}
=\frac{1}{\ell_{Z\mid \GN}(\munderbar{\bz})}\geq 1,\\
A_2>\frac{p_{Z\mid\GN}(\bar{\bz}\mid 1)}{p_{Z\mid\GN}(\bar{\bz}\mid 0)}
=\ell_{Z\mid \GN}(\bar{\bz})\geq 1,
\end{align*}
which yields the following observations:
\begin{itemize}
    \item $\ell_{Z|\GMF}(\bar{\bz})=A_2>1$ and $\ell_{Z|\GN}(\bar{\bz})>1$.
    \item $\ell_{Z|\GMF}(\munderbar{\bz})=1/A_1<1$ and $\ell_{Z|\GN}(\munderbar{\bz})<1$.
    \item For $\bz\ne\munderbar{\bz},\bar{\bz}$, we have  $\ell_{Z|\GMF}(\bz) =\ell_{Z|\GN}(\bz)$.
\end{itemize}
Therefore we have
\begin{align}\label{GammaGNGMF}
\Gamma_{\GN}=\Gamma_{\GMF}.
\end{align}
For any $G\in\calG_Z$, we have
\begin{align}
&R_G(p_{Z|X},\gamma^*_G)\nonumber\\
&=\ofrac{2}\Big(\P(Z\in\Gamma_{G}\mid G=0)+\P(Z\in\Gamma_{G}^c\mid G=1)\Big)\nonumber\\
&\geq\frac{(1-\delta)}{2}\Big(\P(Z\in\Gamma_{G}\mid  \GN=0)\nonumber\\
&\hspace{3cm}+\P(Z\in\Gamma_{G}^c\mid  \GN=1)\Big)\label{ineq:def_uncertainty_set}\\
&\geq\frac{(1-\delta)}{2}\Big(\P(Z\in\Gamma_{\GN}\mid  \GN=0)\nonumber\\
&\hspace{3cm}+\P(Z\in\Gamma_{\GN}^c\mid  \GN=1)\Big)\label{GammaGN}\\
&=\frac{(1-\delta)}{2}\Big(\P(Z\in\Gamma_{\GMF}\mid  \GN=0)\nonumber\\
&\hspace{3cm}+\P(Z\in\Gamma_{\GMF}^c\mid  \GN=1)\Big)\label{eq:def1}\\
&=\ofrac{2}\Big(\P(Z\in\Gamma_{\GMF}\mid  \GMF=0)\nonumber\\
&\hspace{3cm}+\P(Z\in\Gamma_{\GMF}^c\mid  \GMF=1)\Big)\label{eq:def2}\\
&= R_{\GMF}(p_{Z\mid X}, \gamma_{\GMF}^*),\nonumber
\end{align}
where
\begin{itemize}
	\item \cref{ineq:def_uncertainty_set} follows from the definition of $\calG_Z$;
	\item \cref{GammaGN} is because $\Gamma_{\GN}$ is the optimal critical region that minimizes $R_{\GN}(p_{Z\mid X},\cdot)$;
	\item \cref{eq:def1} follows from \cref{GammaGNGMF}; and
	\item \cref{eq:def2} follows from \eqref{MFD}.
\end{itemize}
Therefore, $R_{\GMF}(p_{Z\mid X},\gamma_{\GMF}^*)=\min_{G\in\calG_Z}R_G(p_{Z|X},\gamma^*_G)$. Furthermore, from \cref{GammaGN}, we obtain $R_{\GMF}(p_{Z\mid X},\gamma_{\GMF}^*)=(1-\delta)  R_{\GN}(p_{Z\mid X}, \gamma_{\GN}^*)$, and the theorem is proved.
\end{appendices}

\end{document}